\documentclass[english,aps,preprint,nofootinbib]{revtex4-2}
\usepackage[T1]{fontenc}
\usepackage[latin9]{inputenc}
\usepackage{color}
\usepackage{amsmath}
\usepackage{graphicx}
\usepackage{esint}

\makeatletter

\providecommand{\tabularnewline}{\\}

\usepackage[normalem]{ulem}
\usepackage{xcolor}
\usepackage{babel}
\usepackage{amsfonts}

\makeatother

\usepackage{babel}
\begin{document}
\title{Entanglement Entropy and Quantum Phase Transition in the $O(N)$ $\sigma$-model}
\author{Jiunn-Wei Chen}
\email{jwc@phys.ntu.edu.tw}

\affiliation{Department of Physics and Center for Theoretical Sciences, National
Taiwan University, Taipei 106319, Taiwan}
\affiliation{Leung Center for Cosmology and Particle Astrophysics, National Taiwan
University, Taipei 106319, Taiwan}
\affiliation{Physics Division, National Center for Theoretical Sciences, National
Taiwan University, Taipei 106319, Taiwan}
\affiliation{Center for Theoretical Physics, Massachusetts Institute of Technology,
Cambridge, MA 02139, USA}
\author{Shou-Huang Dai}
\email{shdai@stust.edu.tw}

\affiliation{Center for General Education, Southern Taiwan University of Science
and Technology, Tainan 710301, Taiwan}
\author{Jin-Yi Pang}
\email{axial.pang@gmail.com}

\affiliation{University of Shanghai for Science and Technology, Shanghai 200093,
China}
\affiliation{Department of Physics and Center for Theoretical Sciences, National
Taiwan University, Taipei 106319, Taiwan}
\begin{abstract}
We investigate how entanglement entropy behaves in a non-conformal
scalar field system with a quantum phase transition, by the replica
method. We study the $\sigma$-model in 3+1 dimensions which is $O(N)$
symmetric as the mass squared parameter $\mu^{2}$ is positive, and
undergoes spontaneous symmetry breaking while $\mu^{2}$ becomes negative.
The area law leading divergence of the entanglement entropy is preserved
in both of the symmetric and the broken phases. The spontaneous symmetry
breaking changes the subleading divergence from log to log squared,
due to the cubic interaction on the cone. At the leading order of
the coupling constant expansion, the entanglement entropy reaches
a cusped maximum at the quantum phase transition point $\mu^{2}=0$,
and decreases while $\mu^{2}$ is tuned away from 0 into either phase. 
\end{abstract}
\maketitle

\section{Introduction}

Entanglement\cite{Horodecki2009} is an intriguing feature of Quantum
Mechanics. Its has become increasingly important in quantum information\cite{Vedral2002},
condensed matter\cite{Osborne2002,Vidal2003,Latorre2004,Lin10}, string
theory\cite{Duff2012}, and the interpretation of black hole entropy\cite{Bombelli1986}.
For example, in the condensed matter systems, the ground states of
the conventional superconductors\cite{Bardeen1957,Bardeen1957a} and
the fractional quantum Hall effect\cite{Laughlin1983} are both entangled
states. The level of the entanglement between a certain region and
its surroundings is measured by the entanglement entropy. Such systems
may undergo quantum phase transition at zero temperature by tuning
their physical parameters. The quantum phase transition\cite{Sondhi1997}
is fundamentally different from the conventional thermal phase transition
occurred at non-zero temperature as the former involves a qualitative
change of the ground state of a quantum system. In this paper, we
are particularly interested in the insight about the quantum phase
transition from the point of view of the entanglement entropy.

Traditionally, different phases are classified by the order parameters
keeping track of the symmetries broken in the phase transitions. Valuable
insights on the change of the symmetry and the degrees of freedom
in a phase transition can be obtained by the symmetry breaking order
parameters and thermal dynamical variables, respectively, both of
which can be studied in thermal equilibrium. On the other hand, by
means of the transport coefficients, one can also study how a phase
transition affects the real time responses of the system perturbed
slightly away from thermal equilibrium. One example is the shear viscosity
($\eta$) over the entropy density ($s$). At the phase transition
temperature of a scalar field system, this $\eta/s$ ratio tends to
minimize locally; the minimum has a smooth structure for a crossover,
forms a cusp for a second order phase transition, and has a jump for
a first order phase transition \cite{Chen2008}. This behavior is
seen in many systems in nature and many theoretical models. However,
counter examples can be engineered \cite{Chen:2010vf}.

We aim to pursue analogous understanding on the behavior of the entanglement
entropy under the quantum phase transitions in the present research.
As the entanglement entropy scales with the area of the interface
between two entangled regions, at finite temperatures it becomes subleading
compared with the thermal entropy which scales with the volume of
the system. The quantum phase transition studies involving the entanglement
entropy had focused on the zero temperatures cases, in particular
the lattice systems. For example, previous work on the 1+1 dimensional
transverse Ising model shows the entanglement measure reaches a local
maximum at the second order quantum phase transition \cite{Osborne2002,Lin10}.
Studies on various spin chain models in 1+1 dimensions also reveal
the universal scaling behaviors of the ground-state entanglement near
the quantum critical point \cite{Vidal2003,Latorre2004}. For a broader
review on the entanglement properties in many-body systems, see e.g.
\cite{Amico:2007ag}.

As for the systems of the scalar fields in the continuum, through
the study of the conformal cases \cite{Wilczek2011,Calabrese2004,Calabrese2009,Cardy2014,Holzhey1994,Metlitski2009}
and the non-conformal cases \cite{Calabrese2004,Hertzberg2013}, it
is known that the entanglement entropy ($S_{\mathrm{ent}}$) between
a subregion $A$ and its complement $\bar{A}$ of the spacetime has
a leading behavior exhibiting the power law divergence and the area
law\cite{Srednicki93}, $S_{ent}\sim\Lambda^{d-1}A_{\perp}$, where
$A_{\perp}$ is the area of the $d-1$ dimensional boundary of $A$,
with $d$ denoting the number of spatial dimensions and $\Lambda$
is the momentum cutoff. For the non-conformal scalar field theory
to our particular interest in this paper, the subleading part of the
entanglement entropy, according to \cite{Hertzberg2013} which deals
with a single-component massive scalar field in 3+1 dimensions with
both the cubic and the quartic self-couplings, contains the $\lambda A_{\perp}\Lambda^{2}$
term arising from the quartic interaction $-\frac{\lambda}{4!}\phi^{4}$,
and the $g^{2}A_{\perp}\ln(\Lambda^{-1})$ term from the cubic interaction
$-\frac{g}{3!}\phi^{3}$, where $\lambda$ and $g$ are bare couplings.
They are contributed by the two-loop quantum corrections, and can
be absorbed into the mass renormalization of $\phi$. As a result,
the subleading part of $S_{ent}$, e.g.$\frac{1}{48\pi}\,A_{\perp}m_{r}^{2}\,\ln m_{r}^{2}$,
where $m_{r}$ stands for the renormalized mass of the scalar field,
becomes cutoff-independent, such that as $m_{r}$ varies, the change
in the entanglement entropy is cutoff independent and thus physical.

In this paper, we investigate how the entanglement entropy behaves
in the quantum phase transition of a scalar field system in the continuum
under the spontaneous symmetry breaking (SSB), by the path integral
approach and the replica trick. The theoretical model we explore is
the $O(N)$ $\sigma$-model, which is a weakly coupled $N$-component
scalar field theory, in 3+1 dimensions\footnote{This type of models had been extensively studied on their thermal
phase transition properties, e.g. providing a controlled perturbation
to probe $\eta/s$. See for example \cite{Chen2008,Dobado:2009ek,Dobado:2012zf}.}. Our scalar fields live in a bipartite infinite flat spacetime, where
each of the semi-infinite half-spaces is denoted by $A,\bar{A}$.
In the $O(N)$ symmetric phase, Our model has the quartic interactions
$\frac{\lambda}{4}\left[\sum_{i=1}^{N}(\phi^{i})^{2}\right]^{2}$
as the mass squared $\mu^{2}$ of $\phi^{i}$ is positive. By tuning
$\mu^{2}$ from positive to negative, the $O(N)$ symmetry is spontaneously
broken into $O(N-1)$, and the field composition of the system turns
into a massive mode $\sigma$ with a mass $m_{\sigma}=\sqrt{-2\mu^{2}}$,
and $N-1$ massless Goldstone bosons $\pi^{i}$'s. In the broken phase
the cubic interactions $\frac{1}{\sqrt{2}}gm_{\sigma}\left(\sum_{i=1}^{N-1}(\pi^{i})^{2}\sigma+\sigma^{3}\right)$
emerge, where $g=\sqrt{\lambda}$. This phase is characterized by
the emergence of a non-trivial order, i.e. a non-trivial scalar field
VEV. As the SSB occurs, the system undergoes a quantum phase transition
under which the vacuum states change. We are interested in the leading
and subleading UV divergences of the entanglement entropy in this
paper. We will apply the replica method and the field theory expansion
technique on the cone, to analyze the effect of the quantum phase
transition on the entanglement entropy in the $O(N)$ $\sigma$-model
due to the SSB. This is a perturbatively calculable system due to
weak coupling, and we present our result up to two-loop corrections.
To achieve our goal, we apply some of the approximation technique
from \cite{Hertzberg2013} to the calculation in our symmetric phase.
However, in the broken phase, the calculation method we use is different
from \cite{Hertzberg2013}, and hence the result; this will be explained
later in this Section.

We summarize our result of the entanglement entropy divergence structures
for the $\sigma$ model in the $O(N)$ symmetric and the broken phase
below, according to our renormalization scheme up to two-loop level:
\begin{eqnarray}
S_{\text{ent.}}(\mu^{2},\lambda) & = & \frac{A_{\perp}N}{48\pi}\bigg\{(\ln4)\Lambda^{2}+a\tilde{\lambda}|\mu^{2}|\left(\ln\frac{|\mu^{2}|}{\Lambda^{2}}\right)^{2}\nonumber \\
 &  & \hspace{3cm}+(b+c\tilde{\lambda})|\mu^{2}|\ln\frac{|\mu^{2}|}{\Lambda^{2}}+O\left(\tilde{\lambda}^{2},\frac{|\mu^{2}|}{\Lambda^{2}}\right)\bigg\},\label{4dEE}
\end{eqnarray}
where the coefficients \$a, b, c\$ are given by
\begin{center}
\begin{tabular}{c|c|c}
 & $O(N)$ symmetric  & Spontaneous Symmetry Broken Phase \tabularnewline
\hline 
\hline 
\ \ a\ \  & 0  & $-\frac{3}{(4\pi N)^{2}}\big(N+2\big)$ \tabularnewline
b  & 1  & $\frac{2}{N}$ \tabularnewline
c  & 0  & $-\frac{2}{(4\pi N)^{2}}\big\{9\sqrt{5}\ln\big(\frac{3+\sqrt{5}}{2}\big)+(6\ln2-2)+(3\ln2+2)N\big\}$ \tabularnewline
\end{tabular}. 
\par\end{center}

In these expressions, $N$ is the number of species of the scalar
fields, and $\tilde{\lambda}=\lambda/N$. Note that $\mu$ and $\lambda$
all stand for the renormalized parameters. The area law is clearly
observed, with $A_{\perp}$ denoting the area of the 2-dimensional
boundary surface of $A$. There is also the momentum cutoff $\Lambda$
dependence in the leading divergence.

The $a$, $b$ and $c$ coefficients can be extracted from (where
$\bar{\Lambda}$ is some arbitrary cut-off) 
\begin{eqnarray}
a & = & \frac{\partial^{2}}{\left(\partial\ln|\mu^{2}|\right)^{2}}\frac{\partial}{\partial|\mu^{2}|}\frac{24\pi S_{\text{ent.}}}{A_{\perp}N\tilde{\lambda}}\nonumber \\
b+c\tilde{\lambda} & = & \frac{\partial}{\partial\ln|\mu^{2}|}\frac{\partial}{\partial|\mu^{2}|}\left[\frac{48\pi S_{\text{ent.}}}{A_{\perp}N\tilde{\lambda}}-a|\mu^{2}|\left(\ln\frac{|\mu^{2}|}{\bar{\Lambda}^{2}}\right)^{2}\right].
\end{eqnarray}

(\ref{4dEE}) is obtained by using the renormalized mass, coupling
constant and fields at the tree level, and introduce the counter terms
to cancel the quantum loop corrections, in contrast to \cite{Hertzberg2013}.
The counter terms in the symmetry broken phase and those in the $O(N)$
symmetric phase are of the same origin (c.f. Appendix).

(\ref{4dEE}) shows that the area law is the leading behavior of the
entanglement entropy in both phases, which is expected. In the symmetric
phase, the quartic interactions give rise to the subleading non-analytic
structure $|\mu^{2}|\ln\frac{|\mu^{2}|}{\Lambda^{2}}$ at $O(\tilde{\lambda}^{0})$,
agreeing with the result of \cite{Hertzberg2013}. In the broken phase,
however, a new log squared term $|\mu^{2}|\left(\ln\frac{|\mu^{2}|}{\Lambda^{2}}\right)^{2}$
emerges at $O(\tilde{\lambda})$, which is more divergent than log.
This term is never seen in previous literature, including \cite{Hertzberg2013}.
This term arises from the remnant of cancellation between the two-loop
expansions of the cubic interactions (emerged due to the SSB) and
the counter term. The Goldstone-Goldstone and the Goldstone-massive
mode quartic interactions give rise to the single log divergence (at
$O(\tilde{\lambda})$) and become sub-subleading.

The origin of the log squared term is explained as follows. We find
that the approximation used in \cite{Hertzberg2013} for integrating
the sunset diagrams on the cone due to the cubic interactions are
over-simplified. Careful treatment of those diagrams in our work yields
the double logarithm behavior of the entanglement entropy in the broken
phase. To explain this schematically, \cite{Hertzberg2013} calculates
the Green's function $G_{n}(x,x')$ on the cone by assuming that the
leading divergence is dominated by $x\to x'$, such that the sunset
diagram becomes approximated by the two-loop diagram with one vertex.
The divergence from one of the loops is then canceled by the counter
terms, leaving the single log behavior. However, our analysis shows
that the ``$x$ away from $x'$'' part of the sunset diagrams on
the cone should be taken into account, and as a result the divergence
cannot be canceled by the counter terms, yielding the double log structure
in the broken phase. We refer the readers to Sec. VI for a more in-depth
explanation, and to the Appendix for the detail of the calculation.
We also argue that the log and the double log structures are independent
of the renormalization schemes, as they are of order $\tilde{\lambda}$
quantities, while the influence of different schemes takes effect
at $O(\tilde{\lambda}^{2})$ or higher.

The other difference between our work and \cite{Hertzberg2013} is
in the system setup. \cite{Hertzberg2013} employs a massive single-component
scalar field with $-\frac{g}{3!}\phi^{3}-\frac{\lambda}{4!}\phi^{4}$
interactions. Our model deals with a $N$-component massive scalar
field in the symmetric phase with only the quartic interactions, while
in the broken phase due to SSB, the massless Goldstone bosons and
a massive $\sigma$ field emerge, and the cubic interactions between
them naturally arise beside the quartic ones, in which the massive
scalar mode and the massless Goldstone bosons couple together. In
this sense, our work generalize the result of \cite{Hertzberg2013}.
If we take $N=1$, our model simplifies to \cite{Hertzberg2013}'s
setup, as the massless Goldstone bosons disappear. But in the broken
phase, (\ref{4dEE}) doesn't reduce to \cite{Hertzberg2013}'s result
because we don't take $x\to x'$ approximation for the Green's function
on the cone, as mentioned above.

Moreover, our work also present the novel result in the numerical
behavior of the entanglement entropy of the $O(N)$ $\sigma$-model
under the quantum phase transition due to the SSB. The scaling behavior
of the entanglement entropy thus can be spelled out from (\ref{4dEE}).
(See Sec. \ref{4d} for details.) We find the entanglement entropy
reaches its maximum with a cusp at the transition point $\mu^{2}=0$.
While $|\mu^{2}|$ shifts away from 0 into either phase, the entanglement
entropy decreases as the correlation length reduces away from the
quantum critical point, as shown in Fig. \ref{fig:N4d}. Although
our result of the entanglement entropy change in Fig. \ref{fig:N4d}
is numerical, it would be interesting if an analytic expression of
the universal scaling behavior near the transition point can be unveiled.

In our $O(N)$ sigma model, the phases are classified by the nontrivial
order $\langle\phi^{N}\rangle$. Besides this SSB-driven quantum phase
transitions, there are also phase transitions which do not involve
any ``local'' order parameters. One example is the topologically
ordered phases \cite{Wen90}. In these cases, the entanglement entropy
remains an important quantity identifying the topological order \cite{Jiang12,Li13}.
The entanglement entropy contains a part called the topological entanglement
entropy \cite{Kitaev:2005dm,Levin:2006zz}, which varies in different
topological phases. Such systems, however, are beyond the scope of
this paper.

The structure of this paper is organized as follows. We first review
in Sec. II the fundamentals of the entanglement entropy and the replica
method in quantum field theory. Sec. III presents the entanglement
entropy of $O(N)$ $\sigma$-model in the $O(N)$ symmetric phase,
up to two-loop perturbations. Sec. IV performs analogous calculation
for the broken phase. Sec. V presents the numerical behavior of the
entanglement entropy versus $\mu^{2}$ in both phases. The entanglement
entropy has a cusped maximum at the quantum phase transition point.
Sec. VI. discusses and concludes our work, and presents potential
future applications. The Appendix presents the calculation details
in deriving the divergence structures of the entanglement entropy.

\section{Entanglement entropy and replica method}

The thermal entropy indicates the level of disorder of a system. In
the quantum case, the thermal entropy is given by the von Neumann
entropy 
\begin{equation}
S=\mathrm{Tr}[\rho\ln\rho],\label{eq:entropy}
\end{equation}
where $\rho$ is the density matrix which is normalized to $\mathrm{Tr}\rho=1$.
In the diagonalized basis, the von Neumann entropy reads $S=\sum_{i}p_{i}\ln p_{i}$
, where $p_{i}$ is the probability for each microstates being occupied.
By postulating the occupation of any microstate is equally probable,
(\ref{eq:entropy}) is equivalent to the statistical definition of
the entropy $S\sim\ln\Omega$ up to the Boltzmann constant, reflecting
the total number of accessible microstates in a quantum system of
microcanonical ensemble.

Consider a bipartite system $S$ in a pure state and composed of subsystems
$A$ and $\bar{A}$, where the degrees of freedom in $A$, $\bar{A}$
are entangled in some way. If one is forbidden to access $\bar{A}$,
then for such an observer, $A$ appears in a mixed state, with a reduced
density matrix given by 
\begin{equation}
\rho_{A}=\mathrm{Tr}_{\bar{A}}\rho,
\end{equation}
where $\bar{A}$ is traced out. The information regarding the entanglement
is encoded in $\rho_{A}$. As a result, the level of entanglement
between $A$ and $\bar{A}$ is described by the entanglement entropy,
which is defined by 
\begin{equation}
S_{\text{ent.}}=\mathrm{Tr}_{A}[\rho_{A}\,\ln\rho_{A}].\label{eq:EE_initial}
\end{equation}
Since the vacuum wave-function of $\bar{A}$ is buried in the excited
wave-functions of the ``mix-state'' subsystem $A$ described by
$\rho_{A}$, the expectation value of a local operator can be computed
by 
\begin{equation}
\langle0|\mathcal{O}_{A}|0\rangle=\frac{\text{Tr}[\rho_{A}\,\mathcal{O}_{A}]}{\text{Tr}[\rho_{A}]}.\label{eq:local_expectation}
\end{equation}
One example of such set-up is the black holes. Suppose that the whole
spacetime is in a pure state, but we are unable to access the region
inside the event horizon. Therefore the black hole appears thermal
to an outside observer due to the entanglement between the two regions
separated by the horizon, and so the entropy arise. This is one interpretation
of the black hole entropy.

In this paper, we will consider the case that the system $S$ contains
the whole space, while the subsystems $A$ and $\bar{A}$ each occupies
the infinite half-space, divided by a codimension 2 (with respect
to the whole spacetime) surface. We follow the convention in \cite{Hertzberg2013},
denoting the time $t$ and radial coordinate $x_{\parallel}$ as the
\emph{longitudinal directions}, as they are relevant in our field
theory calculation, while the \emph{transverse directions} indicate
the dimensions of the surface enclosing the subsystem $A$.

In order to calculate the entanglement entropy, we take the generic
scalar field theory as an example and review the replica method in
the following. The entanglement entropy can be calculated by the following
trick\cite{Calabrese2004,Calabrese2009,Hertzberg2013} 
\begin{equation}
S_{\text{ent.}}=-\left.\frac{\partial}{\partial n}\right|_{n\rightarrow1}\ln\,\text{Tr}[\rho_{A}^{n}]=-\mathrm{Tr}[\rho_{A}\,\ln\rho_{A}],\label{eq:replica_trick}
\end{equation}
where the trace is taken within $A$ implicitly. As $n\to1$, we can
take $n=1+\epsilon$ and expand $\ln\mathrm{Tr}[\rho_{A}^{n}]$ in
$\epsilon$ for small $\epsilon$. Then the entanglement entropy can
be spelled out from the $O(\epsilon)$ term.

To calculate (\ref{eq:replica_trick}), we first notice that the elements
of the reduced density matrix $\rho_{A}$ can be expressed in the
path integral formalism, 
\begin{equation}
\langle\varphi_{A}|\rho_{A}|\varphi_{A}^{'}\rangle=\int\mathcal{D}\phi\delta[\phi_{A}(\tau=0^{+})-\varphi_{A}]\delta[\phi_{A}(\tau=0^{-})-\varphi_{A}^{'}]e^{-S_{\text{E}}[\phi=\phi_{A}\oplus\phi_{\bar{A}}]},\label{eq:path_integral_of_reduced_density_matrix}
\end{equation}
where $S_{\text{E}}$ is the action of $\phi$ over the whole Euclidean
space with imaginary time $\tau$. $\phi_{A},\phi_{\bar{A}}$ are
the scalar fields taking values in $A,\bar{A}$ respectively. The
field bases $|\varphi_{A}\rangle$ and $|\varphi_{A}^{'}\rangle$
are states in $A$ at certain time. In this expression, $\bar{A}$
region is traced out. Since taking trace amounts to identifying the
Euclidean time of the initial and the final states, (\ref{eq:path_integral_of_reduced_density_matrix})
implies that $\rho_{A}$ is computed on a manifold where $\bar{A}$
is compactified (in $\tau$ direction) to a cylinder while $A$ is
left open. When we had identified $\phi(\tau=-\infty)=\phi(\tau=\infty)$,
the matrix element of $\rho_{A}^{n}$ is computed on a manifold on
which $\bar{A}$ consists of $n$ cylinders on top of each other while
$A$ becomes a $n$-sheeted spacetime manifold. Taking trace of $\rho_{A}^{n}$
then joins the first sheet with the last for $A$, compactifying it
into a cone with a total angle $2n\pi$ (or an excess angle $\delta=2(n-1)\pi$),
where $n\geq1$. See e.g. Fig. 1 in \cite{Hertzberg2013} or Fig.'s
1 and 2 in \cite{Calabrese2009} for the pictorial realization. As
a result, it is natural to define the trace of $\rho_{A}^{n}$ by
\begin{equation}
\ln\,\mathrm{Tr}[\rho_{A}^{n}]=\ln\left(\frac{Z_{n}}{Z_{1}^{n}}\right),
\end{equation}
where $Z_{n}$ denotes the partition function of the field theory
on the $n$-sheet manifold. ($n=1$ reduces to the case on the ordinary
Euclidean space.) The normalization by $Z_{1}^{n}$ is due to the
requirement that $\left.\mathrm{Tr}[\rho_{A}^{n}]\right|_{n\rightarrow1}=1$.

To summarize, using the replica trick, the entanglement entropy is
calculated by 
\begin{equation}
S_{\text{ent.}}=-\left.\frac{\partial}{\partial n}\left[\ln Z_{n}-n\ln Z_{1}\right]\right|_{n\to1}=-\frac{1}{\epsilon}\left[\ln Z_{n}-n\ln Z_{1}\right].\label{eq:replica_trick_of_path_integral}
\end{equation}
For $n>1$, the replication of sheets takes place in the Euclidean
time coordinate, which belong to the longitudinal directions, while
the transverse directions remain ordinary Euclidean. We will adopt
polar coordinates for the longitudinal part of the spacetime, 
\begin{equation}
(\tau,x_{\parallel})=(r\sin\frac{\theta}{n},r\cos\frac{\theta}{n}),\label{eq:n_sheeted_polar_coordinates}
\end{equation}
where $r\in(0,+\infty)$ and $\theta$ is periodical with $2\pi n$.
Thus the partition function on the $n$-sheet manifold is written
down as 
\begin{equation}
Z_{n}=\int\mathcal{D}\phi\exp\left[-\int d^{d_{\perp}}x_{\perp}\int_{0}^{\infty}rdr\int_{0}^{2\pi n}d\theta\mathcal{L}_{\text{E}}[\phi(r,\theta,x_{\perp})]\right].\label{eq:n_sheeted_partition_function}
\end{equation}
Such expression is valid only for the total spacetime dimensions $d+1>2$.

Since the partition function in quantum field theory is interpreted
as the vacuum energy, and the entanglement entropy of our model is
obtained by (\ref{eq:replica_trick_of_path_integral}), $S_{\mathrm{ent.}}$
can be interpreted as the derivative of the correction to the vacuum
energy due to the cone with respect to the conical deficit angle.
This notion will be more transparent as we calculate the free field
entanglement entropy in the $O(N)$ symmetric phase in next section.

\section{Perturbation expansion of the $O(N)$ $\sigma$-model in the symmetric
phase}

The Euclidean Lagrangian of $3+1$ dimensional $O(N)$ model is given
by 
\begin{equation}
\mathcal{L}_{\text{E}}=\sum_{i=1}^{N}\left[\frac{1}{2}(\partial\phi^{i})^{2}+\frac{1}{2}\mu^{2}(\phi^{i})^{2}\right]+\frac{\lambda}{4}\left[\sum_{i=1}^{N}(\phi^{i})^{2}\right]^{2}+\mathcal{L}_{c.t.},\label{ONLag}
\end{equation}
which has $N$ species of scalar fields with the same mass $\mu$,
admitting $O(N)$ symmetry and quartic interactions. $\mathcal{L}_{c.t.}$
is the counter terms to cancel the loop corrections.

Since in this paper we use the renormalized mass $\mu$ and the renormalized
coupling constant $\lambda$ in the tree level action, the partition
function on $n$-sheet manifold can be expanded with respect to $\lambda$
by 
\begin{eqnarray}
\ln Z_{n} & = & {\color{red}{\color{black}\ln Z_{n,\,0}+\sum_{j=1}^{\infty}\frac{(-\lambda)^{j}}{4^{j}\,j!}\int\left(\prod_{k=1}^{j}d^{d_{\perp}}x_{k\perp}\right)\int_{n}\left(\prod_{k=1}^{j}d^{2}x_{k\parallel}\right)}}\nonumber \\
 &  & {\color{red}\hspace{3cm}{\color{black}\left\{ \langle\left[\sum_{m=1}^{N}\phi^{m}(x_{1})^{2}\right]^{2}...\left[\sum_{n=1}^{N}\phi^{n}(x_{j})^{2}\right]^{2}\rangle_{0}+\mathrm{counter\ terms}\right\} }{\color{black},}}\label{eq:perturbative_expansion}
\end{eqnarray}
where $\ln Z_{n,\,0}$ denotes the $O(\lambda^{0})$ free field part.
The counter terms are introduced to cancel the divergence from the
perturbative corrections of loops, such that the renormalized $\mu$
and $\lambda$ receives no further quantum corrections. $\int_{n}$
is the integral over the 2-dimensional $n$-sheet manifold $\int_{0}^{\infty}rdr\int_{0}^{2\pi n}d\theta$.

\begin{figure}
\includegraphics[scale=0.8]{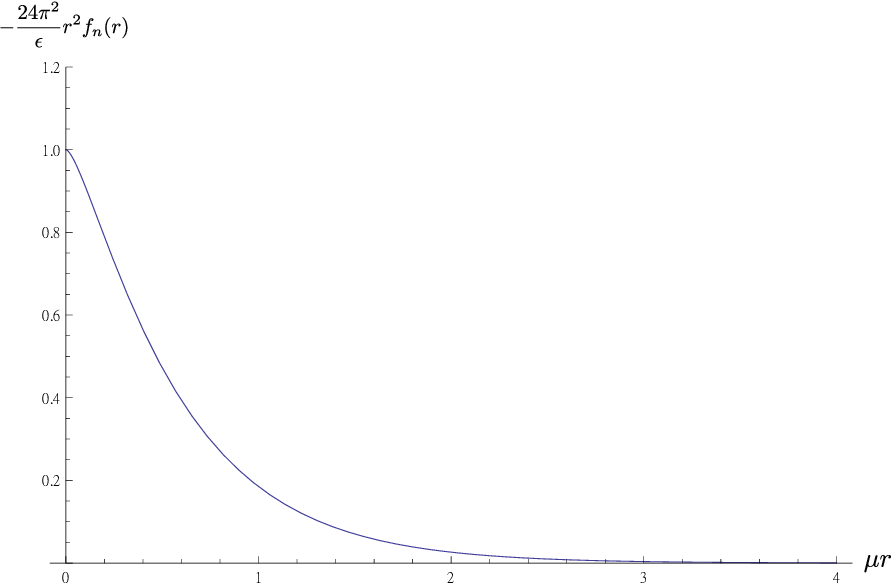} \caption{Behavior of $f_{n}(r)$, giving rise to the subleading divergence
in $S_{\mathrm{ent.}}^{\mathrm{free}}(\mu^{2})$. \label{fig:fn}}
\end{figure}

In the following we calculate the entanglement entropy up to lowest-order
corrections (in this case, $O(\lambda)$ bubble diagrams). The free
field contribution is computed by the following method\cite{Hertzberg2013}.
First, one notices that 
\begin{equation}
\frac{\partial}{\partial\mu^{2}}\ln Z_{n,0}=-\frac{1}{2}\int_{n}d^{d+1}x\,G_{n}(x,x),\label{freeZ}
\end{equation}
where $G_{n}(x,x')$ is the Green's function of the free scalars on
the $n$-sheet Riemann surface, satisfying $(-\nabla^{2}+\mu^{2})\,G_{n}(x,x')=\delta^{d+1}(x-x')$.
The expression for the Green's function on $n$-sheet Riemann surface
$G_{n}(x,x')$ is very complicated, see \cite{Chen:2017mky} for details.
(We will use it later in (\ref{eq:one_loop_correction_of_broken_phase})
in the broken phase.) However, if $x'$ coincides with $x$, however,
the Green's function becomes relatively simpler and we employ an approximation
used in \cite{Calabrese2004} and \cite{Hertzberg2013}. $G_{n}(x,x)$
can be decomposed into 
\begin{equation}
G_{n}(x,x)=G_{1}(0)+f_{n}(r).\label{Gndecomp}
\end{equation}
$G_{1}(0)=G_{1}(|x-x|)$ is the (divergent) Green's function on the
Euclidean flat space, which admits translational invariance, and $f_{n}(r)$
represents the correction to the one-loop vacuum bubble due to the
conical singularity \cite{Hertzberg2013}: 
\begin{eqnarray}
f_{n}(r) & = & \frac{1}{2\pi n}\frac{1-n^{2}}{6n}\int\frac{d^{d_{\perp}}p_{\perp}}{(2\pi)^{d_{\perp}}}K_{0}^{2}(\sqrt{\mu^{2}+p_{\perp}^{2}}r)+\cdots\nonumber \\
 & \stackrel{n=1+\epsilon}{=} & -\frac{\epsilon}{6\pi}\int\frac{d^{d_{\perp}}p_{\perp}}{(2\pi)^{d_{\perp}}}K_{0}^{2}(\sqrt{\mu^{2}+p_{\perp}^{2}}r)+O(\epsilon^{2})+\cdots,\label{fnr}
\end{eqnarray}
with 
\begin{equation}
\int_{0}^{\infty}dy\,y\,K_{0}^{2}(y)=\frac{1}{2}\quad\Rightarrow\quad\int_{0}^{\infty}rdr\,K_{0}^{2}(\sqrt{\mu^{2}+p_{\perp}^{2}}r)=\frac{1}{2}\frac{1}{\mu^{2}+p_{\perp}^{2}},\label{K0int}
\end{equation}
where $K_{0}$ is the modified Bessel function of the second kind
$K_{\nu}$ with $\nu=0$, and $\cdots$ denotes the finite subleading
terms. Since $f_{n}$ is finite for $r>0$ and decays exponentially
at $r\to\infty$ (see Fig. \ref{fig:fn}), one can see that $f_{n}$
vanishes for $n\to1$ (i.e. $\epsilon\to0$, where $\epsilon=n-1$
is the deficit angle of the cone), as expected.

Now we can make use of (\ref{freeZ}) and the subsequent approximation
of $G_{n}(x,x)$ in (\ref{fnr}) and (\ref{K0int}) to calculate the
free fields contribution to the entanglement entropy in the $O(N)$
symmetric phase from (\ref{eq:replica_trick_of_path_integral}): 
\begin{eqnarray*}
S_{\text{ent.}}^{\mathrm{free}}(\mu^{2}) & = & \,-\frac{1}{\epsilon}\left(\ln Z_{1+\epsilon,0}-(1+\epsilon)\ln Z_{1,0}\right)\\
 & = & \frac{N}{\epsilon}\int_{\infty}^{\mu^{2}}\frac{dm^{2}}{2}\int d^{d_{\perp}}x_{\perp}\left\{ \int_{1+\epsilon}d^{2}x_{\parallel}G_{n}(x,x)-(1+\epsilon)\int_{1}d^{2}x_{\parallel}G_{1}(x,x)\right\} .
\end{eqnarray*}
The reason for integrating the mass squared parameter from $\infty$
to $\mu^{2}$ is because we expect the entanglement entropy to vanish
at $\mu^{2}=\infty$, due to vanishing correlation length $\xi\sim\mu^{-1}$.
We can further decompose the integration range into $\int_{\infty}^{\mu^{2}}dm^{2}=\left(\int_{\infty}^{0}+\int_{0}^{\mu^{2}}\right)dm^{2}$,
\begin{eqnarray}
S_{\text{ent.}}^{\mathrm{free}}(\mu^{2}) & = & \frac{N}{\epsilon}\left(\int_{\infty}^{0}+\int_{0}^{\mu^{2}}\right)\frac{dm^{2}}{2}\int d^{d_{\perp}}x_{\perp}\int_{1+\epsilon}d^{2}x_{\parallel}\;f_{1+\epsilon}(r)\label{seefreeeq}\\
 & = & -A_{\perp}\frac{N}{12}\int\frac{d^{d_{\perp}}p_{\perp}}{(2\pi)^{d_{\perp}}}\left[\int_{\infty}^{0}\frac{dm^{2}}{m^{2}+p_{\perp}^{2}}+\int_{0}^{\mu^{2}}\frac{dm^{2}}{m^{2}+p_{\perp}^{2}}\right]\nonumber \\
 & = & -A_{\perp}\frac{N}{12}\int\frac{d^{d_{\perp}}p_{\perp}}{(2\pi)^{d_{\perp}}}\left[\ln\frac{p_{\perp}^{2}}{\Lambda^{2}+p_{\perp}^{2}}+\ln\frac{\mu^{2}+p_{\perp}^{2}}{p_{\perp}^{2}}\right]\nonumber \\
 & = & -A_{\perp}\frac{N}{12}\int\frac{d^{d_{\perp}}p_{\perp}}{(2\pi)^{d_{\perp}}}\ln\frac{\mu^{2}+p_{\perp}^{2}}{\Lambda^{2}+p_{\perp}^{2}},\label{eq:zeroth_order_of_EE}
\end{eqnarray}
where $A_{\perp}=\int d^{d_{\perp}}x_{\perp}$ is the total area of
the transverse space separating $A$ and $\bar{A}$, and $\Lambda$
is the mass scale (or momentum) cutoff in the integration over $dm^{2}$.
In the above calculation, $\int_{n}d^{2}x_{\parallel}G_{1}(0)$ and
$n\int_{1}d^{2}x_{\parallel}G_{1}(0)$ cancel out exactly.


In (\ref{eq:zeroth_order_of_EE}), the leading divergence $\sim\Lambda^{d-1}$
comes from $\int_{\infty}^{0}\frac{dm^{2}}{2}\int d^{d_{\perp}}x_{\perp}\int_{1+\epsilon}d^{2}x_{\parallel}\;f_{1+\epsilon}(r)$.
This can be seen from the $(K_{0})^{2}$ integral in (\ref{fnr})
in 3+1 dimensions ($d=3$): 
\begin{eqnarray}
\frac{1}{\epsilon}\,r^{2}f_{n}(r) & \sim & -\int dp_{\perp}\;p_{\perp}K_{0}^{2}(\sqrt{\mu^{2}+p_{\perp}^{2}}r)=\frac{\mu^{2}}{2}\Big(K_{0}(\mu r)^{2}-K_{1}(\mu r)^{2}\Big)\label{r2fn}\\
 & \sim & \left\{ \begin{array}{l}
-\frac{1}{2}+\frac{(\mu r)^{2}}{2}\Big[(\ln\mu r)^{2}+\cdots\Big]+O\big(\mu^{3}r^{3}\big),\quad\quad r\to0\\
-e^{2\mu r},\hspace{6.37cm}r\to\infty
\end{array}\right.\label{fnexp}
\end{eqnarray}
as displayed in Fig. \ref{fig:fn}. Such behavior means, as the location
of the quantum bubble is far away from the tip of the cone, the effect
of the conical singularity is exponentially suppressed, and the bubble
sees a flat Euclidean space. While the quantum bubble is very close
to the conical point, it is $\frac{1}{r^{2}}$ divergent. Then plug
(\ref{r2fn}) into (\ref{seefreeeq}), one finds that 
\begin{eqnarray}
S_{\text{ent.}}^{\mathrm{free}}\Big|_{\mathrm{leading}} & \sim & \int rdr\int_{\infty}^{0}dm^{2}\frac{m^{2}}{2}\Big(K_{0}(\mu r)^{2}-K_{1}(\mu r)^{2}\Big)\nonumber \\
 & = & \int rdr\;\frac{-1}{3r^{4}}\propto\Lambda^{2}.\label{leadingL}
\end{eqnarray}
On the other hand, the subleading contribution $\sim\ln\Lambda$ to
the entanglement entropy comes from the integral $\int_{0}^{\mu^{2}}\frac{dm^{2}}{2}\int d^{d_{\perp}}x_{\perp}\int_{1+\epsilon}d^{2}x_{\parallel}\;f_{1+\epsilon}(r)$.
It is straightforward to check by substituting the leading term in
(\ref{fnexp}) at small $r$ into (\ref{seefreeeq}), and then going
through the calculation in (\ref{leadingL}).

Since the calculation of $S_{\text{ent.}}^{\mathrm{free}}$ in (\ref{seefreeeq})
stems from (\ref{eq:replica_trick_of_path_integral}), where $f_{n}$
is the correction to the one-loop vacuum bubble due to the cone, it
would be more clear here to comprehend the interpretation of the entanglement
entropy as the derivative of the correction to the vacuum energy due
to the cone with respect to the conical deficit angle, as mentioned
in the end of the previous section.

The perturbation at $O(\lambda)$ level contributes to the partition
function by 
\begin{eqnarray}
\ln Z_{n,1} & = & -\frac{\lambda}{4}\int d^{d_{\perp}}x_{\perp}\int_{n}d^{2}x_{\parallel}\langle\left[\sum_{i=1}^{N}\phi^{i}(x)^{2}\right]^{2}\rangle_{0}\nonumber \\
 & = & -\frac{\lambda}{4}\int d^{d_{\perp}}x_{\perp}\int_{n}d^{2}x_{\parallel}(N^{2}+2N)\left[G_{n}(x,x)^{2}\right],\label{eq:one_loop_correction}
\end{eqnarray}
where we had used $G_{n}^{ij}(x,x^{'})=\delta^{ij}G_{n}(x,x^{'})$.
Then we get 
\begin{eqnarray}
S_{\text{ent.}}(\mu^{2},O(\lambda)) & = & \frac{1}{\epsilon}\frac{\lambda}{4}(N^{2}+2N)\int d^{d_{\perp}}x_{\perp}\left[\int_{1+\epsilon}d^{2}x_{\parallel}G_{1+\epsilon}(x,x)^{2}-(1+\epsilon)\int d^{2}x_{\parallel}G_{1}(x,x)^{2}\right]\nonumber \\
 & = & -\frac{N}{12}\,\lambda\,(N^{2}+2N)\,G_{1}(0)\,A_{\perp}\int\frac{d^{d_{\perp}}p_{\perp}}{(2\pi)^{d_{\perp}}}\frac{1}{\mu^{2}+p_{\perp}^{2}},\label{eq:symmetric_EE_at_one_loop_level}
\end{eqnarray}
in which we have used (\ref{Gndecomp})$\sim$(\ref{K0int}), and
as a result $\int_{n}G_{1}^{2}$ is canceled out by $n\int G_{1}^{2}$,
leaving $\int_{n}G_{1}(0)f_{n}(r)$ as the leading contribution $O(\epsilon)$
in $\epsilon$.

The following counter term is introduced to cancel the above $O(\lambda)$
corrections: 
\begin{equation}
-\frac{1}{2}\delta\mu^{2}\sum_{i=1}^{N}(\phi^{i})^{2},\label{ctON}
\end{equation}
where 
\begin{equation}
\delta\mu^{2}=-(N+2)\lambda G_{1}(0).\label{mrnON}
\end{equation}
This implies the renormalized mass $\mu$ is related to the bare mass
$\mu_{\text{b}}$ by $\mu^{2}=\mu_{\text{b}}^{2}+(N+2)\lambda G_{1}(0)$.
This result is consistent with \cite{Hertzberg2013}. The derivation
of (\ref{mrnON}) is summarized in Appendix for the interested readers.
As a result, the entanglement entropy of the $\sigma$-model in the
$O(N)$ symmetric phase is given by (\ref{eq:zeroth_order_of_EE}).

In 3+1 dimensions, $d_{\perp}=2$, (\ref{eq:zeroth_order_of_EE})
gives rise to the following divergence structure: 
\begin{equation}
S_{\mathrm{ent}}(\mu^{2},\lambda)=\frac{A_{\perp}^{(2)}\Lambda^{2}}{48\pi}N\bigg\{\ln4+\left(\frac{\mu^{2}}{\Lambda^{2}}\right)\ln\left(\frac{\mu^{2}}{\Lambda^{2}}\right)+O\left(\tilde{\lambda}^{2},\frac{\mu^{2}}{\Lambda^{2}}\right)\bigg\},\hspace{0.5cm}\label{4dseesym}
\end{equation}
where $A_{\perp}$ is the area of a 2-dimensional boundary surface
of $A$. All the correction at $O(\lambda)$ are canceled by the counter
terms.

\section{$O(N)$ $\sigma$-model in the symmetry broken phase}

The spontaneous symmetry breaking of $O(N)$ occurs when the mass
squared of the scalar fields $\phi^{i}$ is tuned to $\mu^{2}<0$.
Let's suppose the SSB occurs in the $\phi^{N}$ direction, i.e. $\phi^{N}$
develops a VEV $v$. Then the system is left with $N-1$ massless
Goldstone bosons $\pi^{1},\ldots,\pi^{N-1}$ and 1 massive scalar
$\sigma$, 
\begin{equation}
(\phi^{1},\phi^{2},\ldots,\phi^{N-1},\phi^{N})=(0,0,...,0,v)+(\pi^{1},\pi^{2},...,\pi^{N-1},\sigma),\label{eq:redefinition_of_broken_scalar_field}
\end{equation}
where the condensate $v$ takes the value 
\begin{equation}
v=\langle\phi^{N}\rangle\;=\;\frac{m_{\sigma}}{\sqrt{2}g},\label{eq:field_condensate}
\end{equation}
with $m_{\sigma}=\sqrt{-2\mu^{2}}$ and the new coupling constant
$g=\sqrt{\lambda}$. The Euclidean Lagrangian in the SSB phase becomes
\begin{eqnarray}
\mathcal{L}_{\text{E}} & = & \sum_{i=1}^{N-1}\frac{1}{2}(\partial\pi^{i})^{2}+\frac{1}{2}(\partial\sigma)^{2}+\frac{1}{2}m_{\sigma}^{2}\sigma^{2}\nonumber \\
 &  & +\frac{g}{\sqrt{2}}m_{\sigma}\left(\sum_{i=1}^{N-1}(\pi^{i})^{2}\sigma+\sigma^{3}\right)+\frac{g^{2}}{4}\left(\left[\sum_{i=1}^{N-1}(\pi^{i})^{2}\right]^{2}+\sigma^{4}+2\sum_{i=1}^{N-1}(\pi^{i})^{2}\sigma^{2}\right).\label{eq:broken_sigma_model}
\end{eqnarray}
Compared to the original $O(N)$ $\sigma$-model, the SSB phase contains
not only the quartic interactions but also the cubic ones with coupling
$gm_{\sigma}/\sqrt{2}$.

\begin{figure}
\includegraphics[scale=0.8]{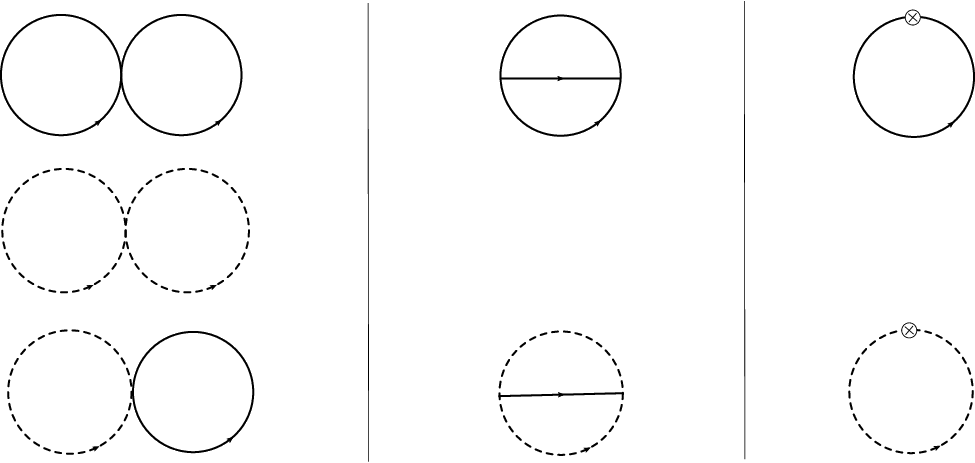} \caption{Feynman diagrams contributing to the two-loop corrections to the entanglement
entropy in (\ref{eq:one_loop_correction_of_broken_phase}). The solid
lines denote the massive mode $\sigma$ while the dashed lines represent
the Goldstone bosons $\pi^{i}$. The left column depicts the quartic
interaction between $\sigma$'s and $\pi^{i}$'s. The middle column
is the two-loops arising from the cubic interactions, while the right
one represent the counter terms' effects to cancel the divergence
due to the diagrams in the left and central column. \label{fig:one_loop_correction}}
\end{figure}

Since we use the renormalized $g$ and $m_{\sigma}$ in the action,
it requires to add counter terms to the classical action to cancel
the loop effects, The partition function up to $O(g^{2})$ corrections
with respect to the new coupling constant $g$ and $m_{\sigma}$ becomes,
\begin{eqnarray}
\ln Z_{n}^{\mathrm{SSB}} & = & \ln Z_{n,0}^{\mathrm{SSB}}-\frac{g^{2}}{4}\int d^{d_{\perp}}x_{\perp}\int_{n}d^{2}x_{\parallel}\left\{ \langle\left[\sum_{i=1}^{N-1}(\pi^{i})^{2}\right]^{2}\rangle+\langle\sigma^{4}\rangle+2\sum_{i=1}^{N-1}\langle(\pi^{i})^{2}\sigma^{2}\rangle\right\} \nonumber \\
 &  & +\frac{g^{2}}{4}m_{\sigma}^{2}\int d^{d_{\perp}}x_{\perp}d^{d_{\perp}}x_{\perp}^{'}\int_{n}d^{2}x_{\parallel}d^{2}x_{\parallel}^{'}\Bigg\{\langle\sigma^{3}(x)\sigma^{3}(x^{'})\rangle\nonumber \\
 &  & \hspace{1cm}\left.\quad+\sum_{i,j=1}^{N-1}\langle[\pi^{i}(x)]^{2}\sigma(x)[\pi^{j}(x^{'})]^{2}\sigma(x^{'})\rangle+2\sum_{i=1}^{N-1}\langle[\pi^{i}(x)]^{2}\sigma(x)\sigma^{3}(x^{'})\rangle\right\} \nonumber \\
 &  & +\mathrm{integral\ of\ counter\ terms}\nonumber \\
 & = & \ln Z_{n,0}^{\mathrm{SSB}}\nonumber \\
 &  & -\frac{g^{2}}{4}\int_{n}d^{d+1}x\left[(N^{2}-1)G_{n}^{\pi}(x,x)^{2}+3G_{n}^{\sigma}(x,x)^{2}+2(N-1)G_{n}^{\pi}(x,x)G_{n}^{\sigma}(x,x)\right]\nonumber \\
 &  & +\frac{g^{2}}{4}m_{\sigma}^{2}\int d^{d_{\perp}}x_{\perp}d^{d_{\perp}}x_{\perp}^{'}\int_{n}d^{2}x_{\parallel}d^{2}x_{\parallel}^{'}\left[6G_{n}^{\sigma}(x',x)^{3}+2(N-1)G_{n}^{\pi}(x',x)^{2}G_{n}^{\sigma}(x',x)\right]\nonumber \\
 &  & +\mathrm{integral\ of\ counter\ terms}.\label{eq:one_loop_correction_of_broken_phase}
\end{eqnarray}
This is up to $O(g^{2})\sim O(\lambda)$. $\ln Z_{1}^{\mathrm{SSB}}$
can also be obtained analogously. Note that the expectation value
here is taken with respect to the new vacuum in the symmetry broken
phase. In terms of Feynman diagrams, these $O(g^{2})\sim O(\lambda)$
terms in the second and the third lines of (\ref{eq:one_loop_correction_of_broken_phase})
are depicted by the one-vertex and two-vertex two-loops in Fig. \ref{fig:one_loop_correction}
respectively. The counter terms added to the action for canceling
the two-loop contributions are 
\begin{equation}
-\frac{g^{2}}{2}\left(\delta^{(1)}\mu^{2}+\frac{m_{\sigma}^{2}}{2}\delta^{(2)}\lambda\right)\sum_{i=1}^{N-1}(\pi^{i}{}^{2})^{2}-\frac{g^{2}}{2}\left(\delta^{(1)}\mu^{2}+\frac{3m_{\sigma}^{2}}{2}\delta^{(2)}\lambda\right)\sigma^{2}-\frac{gm_{\sigma}}{\sqrt{2}}\left(\delta^{(1)}\mu^{2}+\frac{m_{\sigma}^{2}}{2}\delta^{(2)}\lambda\right)\sigma,\label{ctSSB}
\end{equation}
where $\delta^{(1)}\mu^{2}$, $\delta^{(2)}\lambda$ denote the coefficients
of the mass and the coupling constant renormalization counter terms
respectively, with the superscripts $(1),(2)$ labeling the orders
in $\lambda$ (see (\ref{ctterms}) in the Appendix for the meaning
of the superscripts (1), (2)). Note that the wave function renormalization
counter term does not involve up to $O(\lambda)$ here.

\begin{figure}
\includegraphics[scale=0.85]{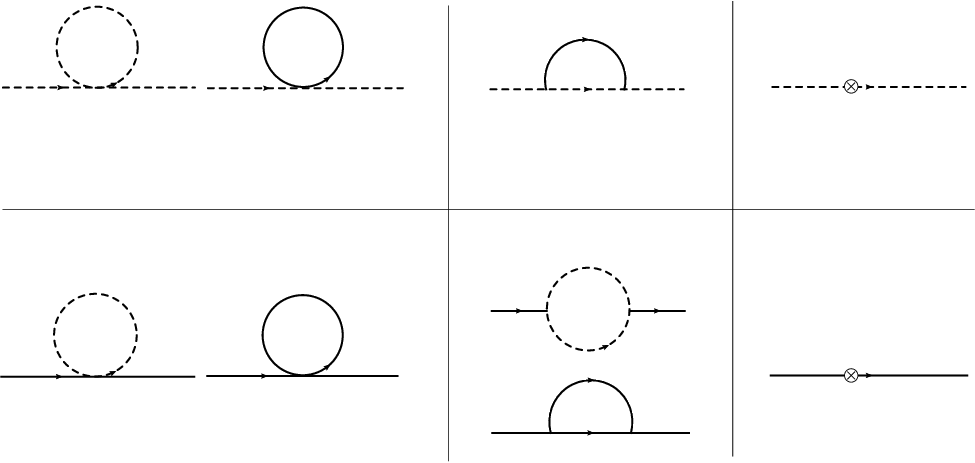} \caption{Feynman diagrams for the mass renormalization of $\pi^{i}$'s (the
upper row) and $\sigma$ (the lower row). The left column represents
the contribution from the quartic interactions, while the middle one
from the cubic interactions. The right column stands for the counter
terms. The solid lines denote $\sigma$ while the dashed lines $\pi^{i}$'s.
\label{fig:mass_shift_and_decay_process}}
\end{figure}

In (\ref{eq:one_loop_correction_of_broken_phase}), we omit the cubic
interactions $\langle\pi^{i}(x)\pi^{i}(x)\sigma(x)\rangle$ and $\langle\sigma(x)\sigma(x)\sigma(x)\rangle$
at $O(g)$ level, because they both vanish. Moreover, the tadpole
diagrams 
\begin{eqnarray}
 &  & \frac{g^{2}m_{\sigma}^{2}}{4}\int_{n}d^{d+1}x\int_{n}d^{d+1}x'\left\{ 9\,G_{n}^{\sigma}(x',x')G_{n}^{\sigma}(x',x)G_{n}^{\sigma}(x,x)\right.\label{tadpoles}\\
 &  & \hspace{1.5cm}\left.+(N-1)^{2}G_{n}^{\pi}(x',x')G_{n}^{\sigma}(x',x)G_{n}^{\pi}(x,x)+6(N-1)\,G_{n}^{\pi}(x',x')G_{n}^{\sigma}(x',x)G_{n}^{\sigma}(x,x)\right\} \nonumber 
\end{eqnarray}
are also dropped out, due to the requirement of the vanishing one-point
function $\langle\sigma\rangle=0$ in the vacuum of broken phase such
that these tadpole corrections to the one-point function $\langle\sigma\rangle$
should be canceled out by the counter terms. This gets rid of the
contribution from (\ref{tadpoles}), and gives rise to the condition
\begin{equation}
\delta^{(1)}\mu^{2}+\frac{m_{\sigma}^{2}}{2}\delta^{(2)}\lambda=-3G_{\sigma}(0)-(N-1)G_{\pi}(0).
\end{equation}
The Feynman diagrams of the non-vanishing two-loop contributions are
displayed in Fig. \ref{fig:mass_shift_and_decay_process}. The explicit
expressions of the counter terms are given by 
\begin{eqnarray}
\delta^{(2)}\lambda & = & 9L_{\sigma}(p^{2}=m_{\sigma}^{2})+(N-1)L_{\pi}(p^{2}=m_{\sigma}^{2}),\nonumber \\
\delta^{(1)}\mu^{2} & = & -3G_{\sigma}(0)-(N-1)G_{\pi}(0)-\frac{9m_{\sigma}^{2}}{2}L_{\sigma}(p^{2}=m_{\sigma}^{2})-\frac{m_{\sigma}^{2}}{2}(N-1)L_{\pi}(p^{2}=m_{\sigma}^{2}),
\end{eqnarray}
where $L_{\pi,\sigma}(p^{2})$ is defined as 
\begin{eqnarray}
L_{\pi,\sigma}(p^{2}) & = & \int\frac{d^{d+1}q}{(2\pi)^{d+1}}G_{\pi,\sigma}(p-q)G_{\pi,\sigma}(q).\label{eqL}
\end{eqnarray}

We can now use (\ref{eq:replica_trick_of_path_integral}) to calculate
the entanglement entropy. The free field contribution is 
\begin{equation}
S_{\text{ent.}}^{\mathrm{SSB(free)}}(m_{\sigma}^{2})=-\frac{N-1}{12}A_{\perp}\int\frac{d^{d_{\perp}}p_{\perp}}{(2\pi)^{d_{\perp}}}\ln\frac{p_{\perp}^{2}}{\Lambda^{2}+p_{\perp}^{2}}-\frac{1}{12}A_{\perp}\int\frac{d^{d_{\perp}}p_{\perp}}{(2\pi)^{d_{\perp}}}\ln\frac{m_{\sigma}^{2}+p_{\perp}^{2}}{\Lambda^{2}+p_{\perp}^{2}}.\label{eq:free_EE_of_broken_phase}
\end{equation}
The first term on the RHS is due to the (massless) Goldstone bosons
$\pi^{1},\ldots,\pi^{N-1}$. The second term is from the massive scalar
$\sigma$.

The two-loop corrections to the entanglement entropy are obtained
as follows. The one-vertex sector (i.e. the second line in (\ref{eq:one_loop_correction_of_broken_phase}),
or the left column of Fig. \ref{fig:one_loop_correction}) is computed
analogously to (\ref{eq:symmetric_EE_at_one_loop_level}), and the
result is 
\begin{eqnarray}
\Delta S_{\text{ent(1-vt)}}^{\text{SSB}}(m_{\sigma}^{2},g^{2}) & = & -A_{\perp}\frac{g^{2}}{12}\Bigg\{\Big(3G_{1}^{\sigma}(0)+(N-1)G_{1}^{\pi}(0)\Big)\int\frac{d^{d_{\perp}}p_{\perp}}{(2\pi)^{d_{\perp}}}\frac{1}{m_{\sigma}^{2}+p_{\perp}^{2}}\nonumber \\
 &  & \hspace{1.8cm}+\Big((N-1)G_{1}^{\sigma}(0)+(N^{2}-1)G_{1}^{\pi}(0)\Big)\int\frac{d^{d_{\perp}}p_{\perp}}{(2\pi)^{d_{\perp}}}\frac{1}{p_{\perp}^{2}}\Bigg\}.\label{SSB1vt}
\end{eqnarray}

As for the two-vertex sector (i.e. the third line in (\ref{eq:one_loop_correction_of_broken_phase}),
or the middle column in Fig. \ref{fig:one_loop_correction}), in general
the Green's function on $n$-sheet Riemann surface $G_{n}(x,x')$
is very complicated. \cite{Hertzberg2013} proposes some approximation
to simplify the calculation, but we discover that part of \cite{Hertzberg2013}'s
approximation is invalid, as explained below. Then we will adopt the
approach used in \cite{Chen:2017mky} instead in our calculation.
The Green's function $G_{n}(x,x')$ on the cone can be decompose into
\begin{equation}
G_{n}(x,x')=G_{1}(|x-x'|)+f_{n}(x,x'),
\end{equation}
where $G_{1}$ represents the $O(\epsilon^{0})$ part while $f_{n}$
the $O(\epsilon)$ (and the higher order) effects of the Green's function,
in analogy to (\ref{Gndecomp}), and $\epsilon$ is the deficit angle
of the cone. It is noticed that $\int_{n}d^{d+1}x\int_{n}d^{d+1}x'G_{1}(|x-x'|)^{3}-n\int_{1}d^{d+1}x\int_{1}d^{d+1}x'G_{1}(|x-x'|)^{3}$
is subleading at UV, where $G_{1}(|x-x'|)$ could either denotes the
Green's function for $\sigma$ or $\pi^{i}$. For convenience we change
the coordinates from $(x,x')$ to $(x,y)$ with $y=x'-x$, such that
the aforementioned subtraction becomes 
\begin{equation}
\int_{n}d^{d+1}x\int_{n}d^{d+1}yG_{1}(|y|)^{3}-n\int_{1}d^{d+1}x\int_{1}d^{d+1}yG_{1}(|y|)^{3}.\label{G1cubic}
\end{equation}

\cite{Hertzberg2013} argues that the divergence of $G_{1}(|y|)$
occurs at small $y\sim\Lambda^{-1}$, i.e. as $x'$ approaches $x$,
where the fields at $x'$ couldn't ``sense'' the existence of the
conical point when $x$ is not close to the conical singularity. This
means that for the divergent part of $G_{1}(|y|)$, the integration
over $\int_{n}d^{d+1}y$ actually takes $2\pi$ angle for $y$ to
encircle $x$ once instead of taking $2n\pi$, and hence contributes
no $n$ factor. The only $n$ factor in the first term of (\ref{G1cubic})
is from $\int_{n}d^{d+1}x$. Then it is straightforward to see that,
at UV, (\ref{G1cubic}) schematically behaves as 
\[
nV\Lambda^{d-1}-nV\Lambda^{d-1},
\]
and the two terms cancel out.

However, we notice that this argument breaks down when $x$ is very
close to the conical singularity, i.e. $|x|<\Lambda^{-1}$, such that
the conical singularity is located within $y<\Lambda^{-1}$. In this
case, both $\int_{n}d^{d+1}x$ and $\int_{n}d^{d+1}y$ give rise to
an $n$ factor, but the former only produce a $A_{\perp}2n\pi/\Lambda^{2}$
coefficient due to restricting $|x|<\Lambda^{-1}$. So in this scenario,
(\ref{G1cubic}) has non-vanishing but finite contribution proportional
to 
\[
A_{\perp}\frac{n^{2}}{\Lambda^{2}}\Lambda^{d-1}-A_{\perp}\frac{n}{\Lambda^{2}}\Lambda^{d-1}=A_{\perp}\Lambda^{d-3}(n^{2}-n),
\]
whose contribution is of higher order in the entanglement entropy.

In terms of $(x,y)$ coordinates, the leading divergent contribution
to $\Delta S_{\text{\textrm{ent}(2-vt)}}^{\text{SSB}}(g^{2})$ comes
from 
\begin{eqnarray}
 &  & \frac{g^{2}m_{\sigma}^{2}}{4}\int_{n}d^{d+1}x\int_{n}d^{d+1}y\:\Big\{18G_{1}^{\sigma}(|y|)^{2}f_{n}^{\sigma}(x,y)\nonumber \\
 &  & \hspace{2.5cm}+2(N-1)\Big[2G_{1}^{\sigma}(|y|)G_{1}^{\pi}(|y|)f_{n}^{\pi}(x,y)+G_{1}^{\pi}(|y|)^{2}f_{n}^{\sigma}(x,y)\Big]\Big\}.\label{2vtleading}
\end{eqnarray}
By further decomposing $y$ into $y_{\perp}$ and $y_{\parallel}$,
the above expression gives rise to the following corrections to the
entanglement entropy: 
\begin{eqnarray}
\Delta S_{\text{ent(2-vt)}}^{\text{SSB}}(m_{\sigma}^{2},g^{2}) & \sim & \frac{g^{2}A_{\perp}}{12}\int\frac{d^{d_{\perp}}p_{\perp}}{(2\pi)^{d_{\perp}}}\frac{m_{\sigma}^{2}}{m_{\sigma}^{2}+p_{\perp}^{2}}\nonumber \\
 &  & \times\left[9\Gamma_{\sigma}(p_{\perp}^{2})+(N-1)\Gamma_{\pi}(p_{\perp}^{2})+2(N-1)\Gamma_{\pi\sigma}(p_{\perp}^{2})\right],\label{eq:ee_2-vt}
\end{eqnarray}
where 
\begin{eqnarray}
\Gamma_{\pi\sigma}(p_{\perp}^{2}) & = & \int\frac{d^{2}k_{\parallel}}{(2\pi)^{2}}\frac{d^{2}k_{\perp}}{(2\pi)^{2}}\frac{1}{k^{2}+m_{\sigma}^{2}}\frac{1}{k_{\parallel}^{2}+(\mathbf{k}_{\perp}+\mathbf{p}_{\perp})^{2}},\nonumber \\
\Gamma_{\pi}(p_{\perp}^{2}) & = & \int\frac{d^{2}k_{\parallel}}{(2\pi)^{2}}\frac{d^{2}k_{\perp}}{(2\pi)^{2}}\frac{1}{k^{2}}\frac{1}{k_{\parallel}^{2}+(\mathbf{k}_{\perp}+\mathbf{p}_{\perp})^{2}},\\
\Gamma_{\sigma}(p_{\perp}^{2}) & = & \int\frac{d^{2}k_{\parallel}}{(2\pi)^{2}}\frac{d^{2}k_{\perp}}{(2\pi)^{2}}\frac{1}{k^{2}+m_{\sigma}^{2}}\frac{1}{k_{\parallel}^{2}+(\mathbf{k}_{\perp}+\mathbf{p}_{\perp})^{2}+m_{\sigma}^{2}}.\nonumber 
\end{eqnarray}
These three expressions correspond to the three one-loop Feynman diagrams
from the top to the bottom in the middle column of Fig. \ref{fig:mass_shift_and_decay_process}
respectively.

One expects that two-loop contributions from (\ref{SSB1vt}) and (\ref{eq:ee_2-vt})
will be canceled by the counter terms in (\ref{ctSSB}). This is indeed
the case for the one-vertex sector (\ref{SSB1vt}); however there
are residual terms from (\ref{eq:ee_2-vt}) after the cancellation,
\begin{eqnarray}
\Delta S_{\text{ent(res)}}^{\text{SSB}}(m_{\sigma}^{2},g^{2}) & \sim & g^{2}\frac{A_{\perp}}{12}\int\frac{d^{2}p_{\perp}}{(2\pi)^{2}}\frac{m_{\sigma}^{2}}{p_{\perp}^{2}+m_{\sigma}^{2}}\left.\left[9D_{\sigma}(p_{\perp}^{2},q^{2})+(N-1)D_{\pi}(p_{\perp}^{2},q^{2})\right]\right|_{q^{2}=m_{\sigma}^{2}}\nonumber \\
 &  & +g^{2}\frac{(N-1)A_{\perp}}{12}\int\frac{d^{2}p_{\perp}}{(2\pi)^{2}}\frac{m_{\sigma}^{2}}{p_{\perp}^{2}+m_{\sigma}^{2}}\left.\left[2D_{\pi\sigma}(p_{\perp}^{2},q^{2})\right]\right|_{q^{2}=0},
\end{eqnarray}
where 
\begin{eqnarray}
D_{\sigma}(p_{\perp}^{2},q^{2}) & = & \Gamma_{\sigma}(p_{\perp}^{2})-L_{\sigma}(q^{2})\neq0,\nonumber \\
D_{\pi}(p_{\perp}^{2},q^{2}) & = & \Gamma_{\pi}(p_{\perp}^{2})-L_{\pi}(q^{2})\neq0,\\
D_{\pi\sigma}(p_{\perp}^{2},q^{2}) & = & \Gamma_{\pi\sigma}(p_{\perp}^{2})-L_{\pi\sigma}(q^{2})\neq0,\nonumber 
\end{eqnarray}
with $L_{\sigma},L_{\pi}$ and $L_{\pi\sigma}$ given by (\ref{eqL}).

The existence of these cancellation remnants means that the expansion
of $S_{\text{ent.}}^{\mathrm{SSB}}$ involves non-trivial subleading
terms. In 3+1 dimensions, the entanglement entropy reads 
\begin{eqnarray}
S_{\text{ent.}}^{\textrm{SSB}}(m_{\sigma}^{2},\lambda) & = & \frac{1}{48\pi}NA_{\perp}\Lambda^{2}\left[\ln4-\frac{\frac{3}{2}N+3}{(4\pi N)^{2}}\tilde{\lambda}\left(\frac{m_{\sigma}^{2}}{\Lambda^{2}}\right)\Big[\ln\left(\frac{m_{\sigma}^{2}}{\Lambda^{2}}\right)\Big]^{2}\right.\label{See4d}\\
 &  & \left.+\left(\frac{1}{N}-\frac{2(N-1)+9\sqrt{5}\,\ln((3+\sqrt{5})/2)}{(4\pi N)^{2}}\tilde{\lambda}\right)\left(\frac{m_{\sigma}^{2}}{\Lambda^{2}}\right)\ln\left(\frac{m_{\sigma}^{2}}{\Lambda^{2}}\right)+O\left(\tilde{\lambda}^{2},\frac{m_{\sigma}^{2}}{\Lambda^{2}}\right)\right].\nonumber 
\end{eqnarray}
The calculation detail for obtaining (\ref{See4d}) from (\ref{2vtleading})
is presented in the Appendix. The $\lambda$-independent part represents
the exact cancellation between the quantum corrections and the counter
term of the quartic sector, leaving only the tree level effects, as
in the $O(N)$ symmetric phase. In this part, the leading divergence
remains the same compared to the symmetric phase, contributed from
$N-1$ $\pi$'s and one $\sigma$, while the $N$-independent log
divergence arises solely from $\sigma$, since it is the only massive
component in the broken phase. The $\lambda$-dependent part in (\ref{See4d})
represents the leftover of the cancellation between the two-vertex
two-loop sector (due to the cubic interactions) and the counter terms,
giving rise to the log squared divergence in the subleading part,
which is more divergent than the subleading log divergence in (\ref{4dseesym})
in the $O(N)$ symmetric phase.

We emphasize that the highest subleading divergence of the entanglement
entropy we obtained in (\ref{See4d}) is log squared in the broken
phase. This is different from the single log result of cubic interaction
in \cite{Hertzberg2013}. The reason is as follows. The author in
\cite{Hertzberg2013} argues that the leading divergence of (\ref{2vtleading})
is contributed by $y\to0$, i.e. both $y_{\perp}\to0$ and $y_{\parallel}\to0$,
such that $f_{n}(x,y)$ becomes (\ref{fnr}), and eventually (\ref{2vtleading})
gives rise to a log divergence as the subleading behavior of entanglement
entropy. But our calculation shows that $y\neq0$ part in $f_{n}(x,y)$
is actually more divergent than log, and yields the log squared term,
by setting $y_{\parallel}\to0$ (which allows $f_{n}(x,y)$ to be
approximated by (\ref{fnr})) while preserving the $y_{\perp}\neq0$
contribution and carrying out the Fourier transformation.

\section{Numerical Results of Entanglement Entropy}

\label{4d}

In this section, we present how the entanglement entropy varies with
mass and how it behaves upon quantum phase transition in the $O(N)$
$\sigma$-model, up to $O(\lambda)$. This result is obtained by the
numerical computation.

\begin{figure}
\includegraphics{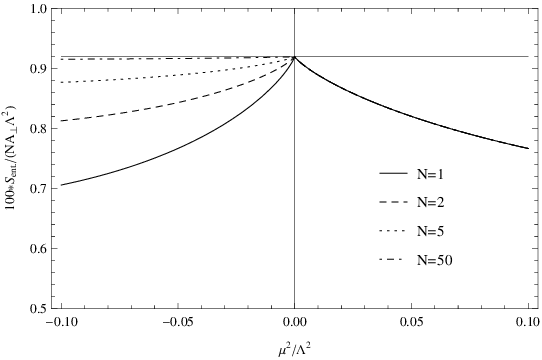} \caption{Entanglement entropy of the $O(N)$ $\sigma$-model in 3+1 dimensions,
in units of $NA_{\perp}\Lambda^{2}$ magnified by a factor of 100,
against the mass squared of fields $\mu^{2}$ normalized by $\Lambda^{2}$.
The left half ($\mu^{2}<0$, with $m_{\sigma}^{2}=-2\mu^{2}$) is
the spontaneous symmetry broken phase while the right half ($\mu^{2}>0$)
is the symmetric phase. In this plot, we take $\tilde{\lambda}=10^{-6}$
in order for the perturbation calculation to be valid. There is a
cusped finite maximum at the quantum phase transition point $\mu^{2}/\Lambda^{2}=0$.
\label{fig:N4d}}
\end{figure}

\begin{figure}
\includegraphics{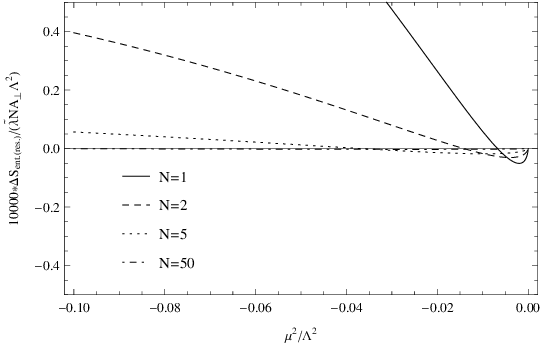} \caption{$\lambda$-dependent subleading part of the entanglement entropy at
$O(\lambda)$ in the symmetry broken phase. Note that this plot is
$\tilde{\lambda}$-independent, as it is in units of $N\tilde{\lambda}A_{\perp}\Lambda^{2}$,
magnified by a factor of 10000. \label{fig:correctionN4d}}
\end{figure}

The numerical value of the entanglement entropy normalized by $A_{\perp}\Lambda^{2}N$
against $\mu^{2}/\Lambda^{2}$ in both phases are plotted in Fig.
\ref{fig:N4d}. The $\mu^{2}>0$ and $\mu^{2}<0$ regions are the
$O(N)$ symmetric phase and the broken phase, respectively. This plot
is shown in terms of renormalized $\mu^{2}$, with $\tilde{\lambda}$
set to $10^{-6}$. In the $O(N)$ symmetric phase, the leading and
subleading parts of $S_{\mathrm{ent.}}$ solely arise from the free
fields, and is $\lambda$-independent. On the other hand, in the symmetry
broken phase, $S_{\mathrm{ent}}^{\mathrm{SSB}}$ contains the free
field contribution and the $\lambda$-dependent remnant from counter
terms cancellation of two-loop corrections due to the cubic interactions.
The latter is shown in Fig. \ref{fig:correctionN4d}.

At the quantum critical point at $\mu^{2}=0$, one expects that the
system acquires scaling symmetry, which gives rise to universal properties,
including the scaling law of the order parameters. For the entanglement
entropy up to the highest subleading term, we found a novel scaling
behavior near the transition point: 
\begin{equation}
S_{\mathrm{ent}}\sim\left\{ \begin{array}{l}
1+\frac{1}{\Lambda^{2}\ln4}\mu^{2}\ln\mu^{2}=(\mu^{2})^{\frac{\mu^{2}}{\Lambda^{2}\ln4}}\hspace{2.55cm}\mbox{(symmetric phase)}
\\
1-\alpha\,m_{\sigma}^{2}\Big(\ln m_{\sigma}^{2}\Big)^{2}=\Big(m_{\sigma}^{2}\Big)^{-\alpha\,m_{\sigma}^{2}\ln m_{\sigma}^{2}}\quad\quad\quad\mbox{(broken phase)},
\end{array}\right.
\end{equation}
where $\alpha$ is a $\tilde{\lambda}$-dependent constant, $\alpha=\frac{\frac{3}{2}N+3}{(4\pi N\Lambda)^{2}\ln4}\tilde{\lambda}$.
Note that $m_{\sigma}^{2}=-2\mu^{2}$ for $\mu^{2}<0$. One immediately
finds that the ``critical exponents'' are not constants as the conventional
quantum critical phenomena suggests; instead they are related to the
mass squared scale itself, and tends to 0 at $\mu^{2}\to0$.

Moreover, one can find in Fig. \ref{fig:N4d} that the entanglement
entropy reduces as $\mu^{2}$ is tuned up. This is because when the
system departs from quantum critical point as $\mu^{2}$ increases
from 0, the correlation length $\xi\sim\mu^{-1}$ decreases, and hence
the level of entanglement reduces. The entanglement entropy has a
finite local maximum with a cusp at the phase transition point $\mu^{2}=0$.
This result is in agreement qualitatively with \cite{Osborne2002}
for the Ising Model.

Such behavior of the entanglement entropy can also be interpreted
from the point of view of lattice models. The spatial derivative term
in action of field theory is regarded as the key for producing non-trivial
entanglement. In the lattice models, the spatial derivative term corresponds
to the difference between the fields at one site and its nearest neighbor,
which is called the lattice link. Without the lattice links, the vacuum
of the total system would be just the direct product of local oscillator's
vacuum at each site, 
\begin{equation}
|\Omega\rangle=|0\rangle_{1}\otimes|0\rangle_{2}\otimes...\,,\label{eq:direct_product}
\end{equation}
i.e. there is no entanglement. However, when the lattice link is present,
vacuum can be the non-trivial superposition of the oscillator's state
at each site. 
\begin{equation}
|\Omega\rangle=\sum_{i_{1}i_{2}...}c_{i_{1}i_{2}...}|\phi_{i_{1}}(1)\rangle\otimes|\phi_{i_{2}}(2)\rangle\otimes...\,,\label{eq:superposition_vacuum}
\end{equation}
where $|\phi_{i_{1}}(1)\rangle$ is the $i$-th state of oscillator
at site $i$.

On the other hand, the mass term in the action corresponds to the
harmonic potential for the oscillators at each site. As the mass increases,
the potential wall becomes more steep, which enhances on-site localization
and suppresses hopping. The tunneling of quantum fluctuations is also
suppressed. All these effects reduce the level of entanglement against
increasing $\mu^{2}$.

When the number of $\phi$ species $N>1$, the Goldstone bosons emerge
in the symmetry broken phase. The contribution to the entanglement
entropy from each Goldstone mode is fixed and independent of $\mu^{2}$,
as been demonstrated in (\ref{eq:free_EE_of_broken_phase}). On the
contrary, the massive mode always has $N=1$ contribution. Therefore
as the mass-varying contribution from $\sigma$ is suppressed by increasing
$N$, and becomes flatter. While $N\rightarrow\infty$, entanglement
entropy in the broken phase is dominated by the massless Goldstone
bosons, and becomes completely flat in the plot.

\section{Discussion, Conclusion and Outlook}

In this paper, we perturbatively calculate the entanglement entropy
of $O(N)$ $\sigma$-model with spontaneous symmetry breaking of $O(N)$
by tuning $\mu^{2}$ from being positive to negative in 3+1 dimensions,
up to the $O(\lambda)$ order. The entanglement entropy of this model
in the symmetric phase is given in (\ref{4dseesym}), while in the
broken phase it becomes (\ref{See4d}). These two expressions are
combined in (\ref{4dEE}). We find that the area law is preserved
in the quantum phase transition. However, due to the emergence of
the cubic interactions, the subleading structure changes from log
in $O(N)$ symmetric phase to log squared in the broken phase. We
also numerically display the behavior of the entanglement entropy
against $\mu^{2}$, as shown in Fig. \ref{fig:N4d}. There occurs
a cusped peak at the quantum phase transition point $\mu^{2}=0$.
While $|\mu^{2}|$ shifts away from 0, the level of entanglement reduces,
in both phases. This implies that such behavior of the entanglement
entropy can be regarded as the signature signifying the quantum phase
transition in the $O(N)$ $\sigma$-model with the order parameter
$\langle\sigma\rangle$.

Generally speaking, in the quantum field theory calculation, each
quantum loop gives rise to a log divergence. The calculation for the
symmetric phase (Section III) shows that the two-loops of $\phi$'s
in general yield the $(\log)^{2}$ divergence, but one of the double
$\log$s is canceled by the counter terms under the on-shell mass
renormalization conditions, leaving the single $\log$ result in (\ref{4dseesym}).
In the broken phase (Section IV), similar cancellation happens for
the two-loop diagrams with one vertex due to the quartic interaction
(in the left column of Fig. \ref{fig:one_loop_correction}) under
the corresponding on-shell renormalization conditions\footnote{see the Appendix for the details.}.
However, the double log divergences of the two-loops with two vertices
at $x$ and $x'$ due to the cubic interactions (in the middle column
of Fig. \ref{fig:one_loop_correction}) can not be canceled by the
counter terms, yielding the $\log$ squared structure in the entanglement
entropy expression in (\ref{See4d}). The double log structure does
not appear in \cite{Hertzberg2013} because their approximation assumes
that the $x\to x'$ part dominates the Green's function on the cone,
which simplifies the sunset diagrams from the cubic interactions in
the middle column of Fig. \ref{fig:one_loop_correction} to the one-vertex
two-loop diagrams in the left column of the same figure. We had argued
that this approximation is incorrect in the last paragraph of Section
IV. If we take into account the \textquotedbl$x$ away from $x'$\textquotedbl{}
contribution, the log squared divergence appears naturally. Note that
the log and the double log structures do not depend on the renormalization
schemes because they are of order $\tilde{\lambda}$ quantities. The
effect of different schemes will take place at $O(\tilde{\lambda}^{2})$
above.

In the expression of the entanglement entropy of our paper, we employ
the renormalized mass and coupling constant in the tree level, such
that the final result is expressed in terms of these renormalized
parameters. From this point of view, our results of the $\log$ divergence
and the coefficients in $S_{\text{ent.}}$ of the $O(N)$ symmetric
phase is consistent with those in \cite{Hertzberg2013} in terms of
bare parameters. Despite that our renormalization prescription is
different from \cite{Hertzberg2013}'s, but they should be equivalent.
This is because their cutoff dependence in the $\lambda$-independent
subleading part is hidden in the renormalized mass, i.e. $m_{r}^{2}=m_{\mathrm{bare}}^{2}+\delta m^{2}(\Lambda)$,
while we use renormalized parameters since the tree level, so that
in our corresponding part the cutoff dependence is manifest. These
two results in fact describe the same physics.

The behavior of the entanglement entropy under spontaneous symmetry
breaking had been studied in some quantum systems, for example \cite{CastroAlvaredo:2011gw,Metlitski:2011pr}.
\cite{CastroAlvaredo:2011gw} shows that, for the bi-partite 1+1 dimensional
Heisenberg ferromagnet with spontaneously broken global symmetry,
the entanglement entropy in general diverges as $\frac{n}{2}\log m$
as $m$ is large, where $m$ is the number of the local degrees of
freedom and is related to the subsystem size $l$ by $m=l^{d}$, and
$n$ is interpreted as the number of zero-energy Goldstone bosons.
In 1+1 dimensions, the boundary between the bipartite subsystems is
a point; as a result, the area law is replaced by a log. In our model,
the subsystem size $l$ is infinitely large, and therefore we don't
have the finite volume ($l^{d}$) contribution in the entanglement
entropy. \cite{Metlitski:2011pr} deals with the $O(N)$ quantum non-linear
sigma models in a finite volume of size L in $d+1\geq2+1$ dimensions.
The $O(N)$ symmetry is spontaneously broken into $O(N-1)$ and the
Goldstone bosons pick up an induced mass by applying a small external
field $\vec{h}$. The entanglement entropy between the bipartite subsystems
is calculated by the wave function method. For the cases of a smooth
boundary in d = 2 and a straight boundary in d = 3 between the subsystems,
the entanglement entropy they obtain has a power law lead term divergent
with the cutoff scale $a$ exhibiting the area law, $A_{\perp}/a^{d-1}$,
and a subleading log term divergent with $L$, $\log(\rho_{s}L/c)$,
where $\rho_{s}$ comes from the Goldstone boson mass due to explicit
symmetry breaking and $c$ is the Goldstone boson velocity. The coefficient
of the log term is proportional to the number of the Goldstone modes.
Their model is the so-called non-linear sigma model as their $\sigma$
field is taken to be infinitely heavy and hence integrated out. In
our model, the Goldstone bosons remain massless so we do not see this
subleading log term, while the massive $\sigma$ field in our model
is included, giving rise to our double log term (due to the cubic
interaction) and log term (due to the quartic interaction) in the
broken phase. The absence of the massive $\sigma$ field in \cite{Metlitski:2011pr}
also explains why their entanglement entropy lacks our double log
term.

There are many interesting directions to be explored based on this
work. Straight forward generalization includes introducing the gauge
fields into the $O(N)$ $\sigma$-model, imposing a chemical potential,
or external electric-magnetic fields, and then study the quantum phase
transition in terms of entanglement. Such a system would be more complicated
yet more realistic. Moreover, in high energy physics, pion gas with
isospin chemical potential has quantum phase transition into Bose-Einstein
condensate, which is also of interest to study from the perspective
of entanglement entropy.


\begin{acknowledgments}
The authors would like to thank Feng-Li Lin, Qun Wang and Shi Pu for
helpful discussions. This work is supported by the MOST, NTU-CTS and
the NTU-CASTS of Taiwan. JWC is partly supported by the Ministry of
Science and Technology, Taiwan, under Grant No. 108-2112-M-002-003-MY3
and the Kenda Foundation. 
\end{acknowledgments}


\appendix

\section*{Appendix: \ \ Renormalization of O(N) model and Entanglement Entropy}

In the symmetric phase, the action of O(N) sigma-model is composed
of the renormalized part and the counter terms, 
\begin{eqnarray}
\mathcal{L} & = & \sum_{i=1}^{N}\left[\frac{1}{2}\left(\partial\phi^{i}\right)^{2}-\frac{1}{2}\mu^{2}\left(\phi^{i}\right)^{2}\right]-\frac{\lambda}{4}\left[\sum_{i=1}^{N}\left(\phi^{i}\right)^{2}\right]^{2}\nonumber \\
 &  & +\sum_{i=1}^{N}\left[\frac{1}{2}\delta Z\left(\partial\phi^{i}\right)^{2}-\frac{1}{2}\delta\mu^{2}\left(\phi^{i}\right)^{2}\right]-\frac{\delta\lambda}{4}\left[\sum_{i=1}^{N}\left(\phi^{i}\right)^{2}\right]^{2},\label{eq:L_with_ct}
\end{eqnarray}
where 
\begin{eqnarray}
\delta Z & = & O(\lambda^{2}),\nonumber \\
\delta\mu^{2} & = & \lambda\,\delta^{(1)}\mu^{2}+O(\lambda^{2}),\label{ctterms}\\
\delta\lambda & = & \lambda^{2}\,\delta^{(2)}\lambda+O(\lambda^{3})\nonumber 
\end{eqnarray}
denote the wave-function, mass, and coupling constant counter terms
expanded w.r.t. $\lambda$. The superscripts $(1),(2)$ in $\delta^{(1)}\mu^{2}$
and $\delta^{(2)}\lambda$ label the coefficients of the corresponding
$\lambda$ expansion order of $\delta\mu^{2}$ and $\delta\lambda$
respectively. It can be demonstrated that wave-function renormalization
counter term $\delta Z$ and the coupling renormalization counter
term are no less than second order. In $O(\lambda)$, only the mass
renormalization counter term $\delta\mu^{2}$ is involved. So we have,
\begin{eqnarray}
\mathcal{L} & = & \sum_{i=1}^{N}\left[\frac{1}{2}\left(\partial\phi^{i}\right)^{2}-\frac{1}{2}\mu^{2}\left(\phi^{i}\right)^{2}\right]-\frac{\lambda}{4}\left[\sum_{i=1}^{N}\left(\phi^{i}\right)^{2}\right]^{2}\nonumber \\
 &  & -\sum_{i=1}^{N}\left[\frac{1}{2}\delta\mu^{2}\left(\phi^{i}\right)^{2}\right]
\end{eqnarray}
up to $O(\lambda)$. After the SSB, however, the fields split into
\begin{eqnarray}
(\phi^{1},\phi^{2},...,\phi^{N-1},\phi^{N}) & = & (0,0,...,0,\frac{m_{\sigma}}{\sqrt{2\lambda}})+(\pi^{1},\pi^{2},...,\pi^{N-1},\sigma).
\end{eqnarray}
where $m_{\sigma}=\sqrt{-2\mu^{2}}$, and the Lagrangian with the
counter terms in (\ref{eq:L_with_ct}) becomes 
\begin{eqnarray}
\mathcal{L} & = & \sum_{i=1}^{N-1}\left[\frac{1}{2}\left(\partial\pi^{i}\right)\right]+\left[\frac{1}{2}\left(\partial\sigma\right)^{2}-\frac{1}{2}m_{\sigma}^{2}\sigma^{2}\right]\nonumber \\
 &  & -\frac{\lambda}{4}\left[\sum_{i=1}^{N-1}\left(\pi^{i}\right)^{2}\right]^{2}-\frac{\lambda}{4}\sigma^{4}-\frac{\lambda}{2}\left[\sum_{i=1}^{N-1}\left(\pi^{i}\right)^{2}\right]\sigma^{2}\nonumber \\
 &  & -\sqrt{\frac{\lambda}{2}}m_{\sigma}\left[\sum_{i=1}^{N-1}\left(\pi^{i}\right)^{2}\right]\sigma-\sqrt{\frac{\lambda}{2}}m_{\sigma}\sigma^{3}\nonumber \\
 &  & -\frac{\lambda}{2}\left[\delta^{(1)}\mu^{2}+\frac{1}{2}m_{\sigma}^{2}\delta^{(2)}\lambda\right]\sum_{i=1}^{N-1}\left[\left(\pi^{i}\right)^{2}\right]\nonumber \\
 &  & -\frac{\lambda}{2}\left[\delta^{(1)}\mu^{2}+\frac{3}{2}m_{\sigma}^{2}\delta^{(2)}\lambda\right]\sigma^{2}\nonumber \\
 &  & -\sqrt{\frac{\lambda}{2}}m_{\sigma}\left[\delta^{(1)}\mu^{2}+\frac{1}{2}m_{\sigma}^{2}\delta^{(2)}\lambda\right]\sigma+O(\lambda^{2}).\label{eq:L_with_ct_SSB}
\end{eqnarray}
up to $O(\lambda)$. By demanding the 1-loop mass correction be zero
(see (\ref{eq:M_ren_SSB}) below) and the tadpole contribution (\ref{eq:tadpole_vanish})
vanish, the two different combinations of the coefficients in the
counter terms in the last three lines of (\ref{eq:L_with_ct_SSB})
are fixed, and the infinity in this Lagrangian is canceled. The calculation
is in the following.

The renormalization scheme depends on the renormalization condition.
To fix the mass renormalization counter terms which appear in both
the symmetric and the broken phases, we choose the on-shell mass renormalization
condition, such that the renormalized masses of $\phi$ (in the symmetric
phase) and $\sigma$ (in the broken phase) equal to their tree level
ones respectively. Assuming the propagator takes the form $G^{-1}(p^{2})=p^{2}+M^{2}(p^{2})$.

As a result, in the symmetric phase, $\phi$ physical mass is defined
by the renormalization condition 
\begin{equation}
M_{\phi}^{2}(p^{2}=\mu^{2})=\mu^{2},
\end{equation}
and in the broken phase, $\sigma$ physical mass is fixed by 
\begin{equation}
M_{\sigma}^{2}(p^{2}=m_{\sigma}^{2})=m_{\sigma}^{2}=-2\mu^{2}.\label{eq:M_ren_SSB}
\end{equation}

In the $O(N)$ symmetric phase, the entanglement entropy of the scalar
field theory in $3+1$ dimensions is UV divergent, 
\begin{eqnarray}
S_{\text{ent.}}(\mu^{2},\lambda) & = & \#A_{\perp}\Lambda^{2}\left[\left(a_{0}+a_{1}\lambda\right)\right.\nonumber \\
 &  & \left.+\left(b_{0}+b_{1}\lambda\right)\left(\frac{|\mu^{2}|}{\Lambda^{2}}\right)\ln\left(\frac{|\mu^{2}|}{\Lambda^{2}}\right)+O\left(\lambda^{2},\frac{|\mu^{2}|}{\Lambda^{2}}\right)\right],
\end{eqnarray}
where $\Lambda$ is cut-off of momentum. This is indeed the case for
the $O(N)$ $\sigma$-model in symmetric phase. As $|\mu^{2}|/\Lambda^{2}\rightarrow0$,
the leading order is $a_{0}+a_{1}\lambda+O(\lambda^{2})$, while the
subleading order is 
\begin{eqnarray*}
\left(b_{0}+b_{1}\lambda+O(\lambda^{2})\right)\left(\frac{|\mu^{2}|}{\Lambda^{2}}\right)\ln\left(\frac{|\mu^{2}|}{\Lambda^{2}}\right).
\end{eqnarray*}
In this phase, the wave-function and the coupling renormalization
counter terms are not involved, so the renormalization condition at
$O(\lambda)$ level is given by 
\begin{eqnarray}
M_{\phi,(0)}^{2}+M_{\phi,(1)}^{2}\lambda+O(\lambda^{2}) & = & \mu^{2},
\end{eqnarray}
where 
\begin{eqnarray}
M_{\phi,(0)}^{2} & = & \mu^{2},\\
M_{\phi,(1)}^{2} & = & N\left[\delta^{(1)}\mu^{2}+(N+2)G_{\phi}(0)\right].
\end{eqnarray}
Thus, we have 
\begin{eqnarray}
\delta\mu^{2} & = & -\lambda(N+2)G_{\phi}(0)+O(\lambda^{2}).
\end{eqnarray}

The $S_{\text{ent.}}$ can be expressed by the transverse mass $m_{\perp}^{2}$,
\begin{eqnarray}
S_{\text{ent.}}(\mu^{2}>0) & = & -\frac{NA_{\perp}}{12}\int\frac{d^{2}p_{\perp}}{(2\pi)^{2}}\ln\frac{p_{\perp}^{2}+m_{\perp}^{2}(\mu^{2})}{p_{\perp}^{2}+\Lambda^{2}}.
\end{eqnarray}
With the counter term, the transverse mass is just renormalized mass,
\begin{eqnarray}
m_{\perp}^{2} & = & \mu^{2}+\lambda(N+2)G_{\phi}(0)+\lambda\delta^{(1)}\mu^{2}+O(\lambda^{2})\nonumber \\
 & = & \mu^{2}+O(\lambda^{2}).
\end{eqnarray}
If we express the entanglement entropy in the following general form,
\begin{equation}
S_{\text{ent.}}(\mu^{2}>0)=\#A_{\perp}\Lambda^{2}\left[\left(a_{0}+a_{1}\lambda\right)+\left(b_{0}+b_{1}\lambda\right)\left(\frac{\mu^{2}}{\Lambda^{2}}\right)\ln\left(\frac{\mu^{2}}{\Lambda^{2}}\right)+O\left(\lambda^{2},\frac{\mu^{2}}{\Lambda^{2}}\right)\right],
\end{equation}
the coefficients then are 
\begin{eqnarray}
a_{0} & = & N\log4,\nonumber \\
b_{0} & = & N,\\
a_{1} & = & b_{1}\;=\;0.\nonumber 
\end{eqnarray}

In the broken phase, to fix the coupling constant renormalization
counter terms which now appear at the leading order, we use the on-shell
mass renormalization condition (\ref{eq:M_ren_SSB}) along with the
requirement that the tadpole contribution of $\sigma$ vanishes, 
\begin{equation}
\langle\sigma(x)\rangle=V_{\sigma}G_{\sigma}(x)=0,\label{eq:tadpole_vanish}
\end{equation}
where the tadpole VEV is decomposed into the coefficient $V_{\sigma}$
and the Green's function $G_{\sigma}(x)$. They give rise to the conditions
\begin{eqnarray}
M_{\sigma,(0)}^{2}+M_{\sigma,(1)}^{2}(p^{2}=m_{\sigma}^{2})\lambda+O(\lambda^{2}) & = & m_{\sigma}^{2},\\
V_{\sigma,(0)}+V_{\sigma,(1)}\sqrt{\lambda}+O(\lambda) & = & 0,
\end{eqnarray}
where the subscripts $(1),(2)$ denotes the coefficients of the corresponding
order in $\lambda$ expansions. We find 
\begin{eqnarray}
M_{\sigma,(0)}^{2} & = & m_{\sigma}^{2},\\
M_{\sigma,(1)}^{2}(p^{2}=m_{\sigma}^{2}) & = & 3G_{\sigma}(0)+(N-1)G_{\pi}(0)\nonumber \\
 &  & -9m_{\sigma}^{2}L_{\sigma}(p^{2}=m_{\sigma}^{2})-(N-1)m_{\sigma}^{2}L_{\pi}(p^{2}=m_{\sigma}^{2})\nonumber \\
 &  & +\left(\delta^{(1)}\mu^{2}+\frac{3}{2}m_{\sigma}^{2}\delta^{(2)}\lambda\right),\\
V_{\sigma,(0)} & = & 0,\\
V_{\sigma,(1)} & = & -\frac{m_{\sigma}}{\sqrt{2}}\left[3G_{\sigma}(0)+(N-1)G_{\pi}(0)\right.\nonumber \\
 &  & \left.+\left(\delta^{(1)}\mu^{2}+\frac{1}{2}m_{\sigma}^{2}\delta^{(2)}\lambda\right)\right],
\end{eqnarray}
in which $L(p^{2})$ is defined by 
\begin{eqnarray}
L(p^{2}) & = & \int\frac{d^{4}q}{(2\pi)^{4}}G(p-q)G(q).
\end{eqnarray}
Thus, we have 
\begin{eqnarray}
\delta\mu^{2} & = & -\lambda\left[3G_{\sigma}(0)+(N-1)G_{\pi}(0)\right.\nonumber \\
 &  & \left.+\frac{9}{2}m_{\sigma}^{2}L_{\sigma}(p^{2}=m_{\sigma}^{2})+\frac{1}{2}(N-1)m_{\sigma}^{2}L_{\pi}(p^{2}=m_{\sigma}^{2})\right]+O(\lambda^{2}),\\
\delta\lambda & = & \lambda^{2}\left[9L_{\sigma}(p^{2}=m_{\sigma}^{2})+(N-1)L_{\pi}(p^{2}=m_{\sigma}^{2})\right]+O(\lambda^{3}).
\end{eqnarray}

Now the entanglement entropy reads 
\begin{eqnarray}
S_{\text{ent.}}(\mu^{2}<0) & = & -\frac{A_{\perp}}{12}\int\frac{d^{2}p_{\perp}}{(2\pi)^{2}}\ln\frac{p_{\perp}^{2}+m_{\sigma,\perp}^{2}(m_{\sigma}^{2})}{p_{\perp}^{2}+\Lambda^{2}}\nonumber \\
 &  & -\frac{(N-1)A_{\perp}}{12}\int\frac{d^{2}p_{\perp}}{(2\pi)^{2}}\ln\frac{p_{\perp}^{2}+m_{\pi,\perp}^{2}(m_{\sigma}^{2})}{p_{\perp}^{2}+\Lambda^{2}}\nonumber \\
 &  & +\lambda\frac{A_{\perp}}{12}\int\frac{d^{2}p_{\perp}}{(2\pi)^{2}}\frac{m_{\sigma}^{2}}{p_{\perp}^{2}+m_{\sigma}^{2}}\left.\left[D_{\sigma}(p_{\perp}^{2},q^{2})+(N-1)D_{\pi}(p_{\perp}^{2},q^{2})\right]\right|_{q^{2}=m_{\sigma}^{2}}\nonumber \\
 &  & +\lambda\frac{(N-1)A_{\perp}}{12}\int\frac{d^{2}p_{\perp}}{(2\pi)^{2}}\frac{m_{\sigma}^{2}}{p_{\perp}^{2}+m_{\sigma}^{2}}\left.\left[D_{\pi\sigma}(p_{\perp}^{2},q^{2})\right]\right|_{q^{2}=0}+O(\lambda^{2}),\label{eq:EE_SSB}
\end{eqnarray}
where 
\begin{eqnarray}
D_{\sigma}(p_{\perp}^{2},q^{2}) & = & 9\int\frac{d^{2}k_{\parallel}}{(2\pi)^{2}}\frac{d^{2}k_{\perp}}{(2\pi)^{2}}\frac{1}{k^{2}+m_{\sigma}^{2}}\left[\frac{1}{k_{\parallel}^{2}+(\mathbf{k}_{\perp}+\mathbf{p}_{\perp})^{2}+m_{\sigma}^{2}}-\frac{1}{(\mathbf{k}+\mathbf{q})^{2}+m_{\sigma}^{2}}\right]\nonumber \\
 & = & 9\left[\Gamma_{\sigma}(p_{\perp}^{2})-L_{\sigma}(q^{2})\right],\\
D_{\pi}(p_{\perp}^{2},q^{2}) & = & \int\frac{d^{2}k_{\parallel}}{(2\pi)^{2}}\frac{d^{2}k_{\perp}}{(2\pi)^{2}}\frac{1}{k^{2}}\left[\frac{1}{k_{\parallel}^{2}+(\mathbf{k}_{\perp}+\mathbf{p}_{\perp})^{2}}-\frac{1}{(\mathbf{k}+\mathbf{q})^{2}}\right]\nonumber \\
 & = & \Gamma_{\pi}(p_{\perp}^{2})-L_{\pi}(q^{2}),\\
D_{\pi\sigma}(p_{\perp}^{2},q^{2}) & = & 2\int\frac{d^{2}k_{\parallel}}{(2\pi)^{2}}\frac{d^{2}k_{\perp}}{(2\pi)^{2}}\frac{1}{k^{2}+m_{\sigma}^{2}}\left[\frac{1}{k_{\parallel}^{2}+(\mathbf{k}_{\perp}+\mathbf{p}_{\perp})^{2}}-\frac{1}{(\mathbf{k}+\mathbf{q})^{2}}\right]\nonumber \\
 & = & 2\left[\Gamma_{\pi\sigma}(p_{\perp}^{2})-L_{\pi\sigma}(q^{2})\right].
\end{eqnarray}
The explicit calculation of $D$ is shows that

\begin{eqnarray}
D_{\sigma}(p_{\perp}^{2},q^{2}) & = & \int\frac{d^{2}k_{\parallel}}{(2\pi)^{2}}\frac{d^{2}k_{\perp}}{(2\pi)^{2}}\frac{9}{k^{2}+m_{\sigma}^{2}}\left[\frac{1}{k^{2}+m_{\sigma}^{2}+2\mathbf{k}_{\perp}\cdot\mathbf{p}_{\perp}+p_{\perp}^{2}}-\frac{1}{k^{2}+m_{\sigma}^{2}+2\mathbf{k}\cdot\mathbf{q}+q^{2}}\right]\nonumber \\
 & = & 9\int\frac{d^{4}l}{(2\pi)^{4}}\int_{0}^{1}dx\left[\left(\frac{1}{l^{2}+m_{\sigma}^{2}-x^{2}p_{\perp}^{2}+xp_{\perp}^{2}}\right)^{2}-\left(\frac{1}{l^{2}+m_{\sigma}^{2}-x^{2}q^{2}+xq^{2}}\right)^{2}\right]\nonumber \\
 & = & \frac{9}{(4\pi)^{2}}\int_{0}^{1}dx\ln\left(\frac{m_{\sigma}^{2}+x(1-x)q^{2}}{m_{\sigma}^{2}+x(1-x)p_{\perp}^{2}}\right),\label{eq:dsigma}
\end{eqnarray}

\begin{eqnarray}
D_{\pi}(p_{\perp}^{2},q^{2}) & = & \int\frac{d^{2}k_{\parallel}}{(2\pi)^{2}}\frac{d^{2}k_{\perp}}{(2\pi)^{2}}\frac{1}{k^{2}}\left[\frac{1}{k^{2}+2\mathbf{k}_{\perp}\cdot\mathbf{p}_{\perp}+p_{\perp}^{2}}-\frac{1}{k^{2}+2\mathbf{k}\cdot\mathbf{q}+q^{2}}\right]\nonumber \\
 & = & \int\frac{d^{4}l}{(2\pi)^{4}}\int_{0}^{1}dx\left[\left(\frac{1}{l^{2}-x^{2}p_{\perp}^{2}+xp_{\perp}^{2}}\right)^{2}-\left(\frac{1}{l^{2}-x^{2}q^{2}+xq^{2}}\right)^{2}\right]\nonumber \\
 & = & \frac{1}{(4\pi)^{2}}\ln\left(\frac{q^{2}}{p_{\perp}^{2}}\right),\label{eq:DEQ_83}
\end{eqnarray}

\begin{eqnarray}
D_{\pi\sigma}(p_{\perp}^{2},q^{2}) & = & 2\int\frac{d^{2}k_{\parallel}}{(2\pi)^{2}}\frac{d^{2}k_{\perp}}{(2\pi)^{2}}\frac{1}{k^{2}+m_{\sigma}^{2}}\left[\frac{1}{k^{2}+2\mathbf{k}_{\perp}\cdot\mathbf{p}_{\perp}+p_{\perp}^{2}}-\frac{1}{k^{2}+2\mathbf{k}\cdot\mathbf{q}+q^{2}}\right]\nonumber \\
 & = & 2\int\frac{d^{4}l}{(2\pi)^{4}}\int_{0}^{1}dx\left[\left(\frac{1}{l^{2}+(1-x)m_{\sigma}^{2}-x^{2}p_{\perp}^{2}+xp_{\perp}^{2}}\right)^{2}\right.\nonumber \\
 &  & \hspace{5cm}\left.-\left(\frac{1}{l^{2}+(1-x)m_{\sigma}^{2}-x^{2}q^{2}+xq^{2}}\right)^{2}\right]\nonumber \\
 & = & \frac{2}{(4\pi)^{2}}\int_{0}^{1}dx\ln\left[\frac{m_{\sigma}^{2}+xq^{2}}{m_{\sigma}^{2}+xp_{\perp}^{2}}\right].\label{eq:DEQ_84}
\end{eqnarray}

The transverse masses of $\pi$ and $\sigma$ are 
\begin{eqnarray}
m_{\sigma,\perp}^{2} & = & m_{\sigma}^{2}+\lambda\Big[3G_{\sigma}(0)+(N-1)G_{\pi}(0)-9m_{\sigma}^{2}L_{\sigma}(p^{2}=m_{\sigma}^{2})-(N-1)m_{\sigma}^{2}L_{\pi}(p^{2}=m_{\sigma}^{2})\Big]\nonumber \\
 &  & \quad\;+\lambda\left[\delta^{(1)}\mu^{2}+\frac{3}{2}m_{\sigma}^{2}\delta^{(2)}\lambda\right]+O(\lambda^{2}),\\
m_{\pi,\perp}^{2} & = & \lambda\left[G_{\sigma}(0)+(N+1)G_{\pi}(0)-2m_{\sigma}^{2}L_{\pi\sigma}(p^{2}=0)\right]\nonumber \\
 &  & +\lambda\left[\delta^{(1)}\mu^{2}+\frac{1}{2}m_{\sigma}^{2}\delta^{(2)}\lambda\right]+O(\lambda^{2}).
\end{eqnarray}
The mass of $\pi$ is always zero. This gives rise to the condition
\begin{eqnarray}
\delta^{(1)}\mu^{2}+\frac{1}{2}m_{\sigma}^{2}\delta^{(2)}\lambda & = & -3G_{\sigma}(0)-(N-1)G_{\pi}(0).
\end{eqnarray}
One can demonstrate that 
\begin{eqnarray}
M_{\pi}^{2}(p^{2}=0) & = & 0+O(\lambda^{2}),\\
m_{\pi,\perp}^{2} & = & 0+O(\lambda^{2}).
\end{eqnarray}
And, by normalizing $m_{\sigma}$ at $q^{2}=m_{\sigma}^{2}$, we have
\begin{eqnarray}
m_{\sigma,\perp}^{2} & = & m_{\sigma}^{2}+O(\lambda^{2}).
\end{eqnarray}

The explicit calculation of residual part of the entanglement entropy
that cannot be absorbed by mass renormalization shows that $\Delta S_{\text{ent.}}^{\text{res.}}$
is given by 
\begin{eqnarray*}
\Delta S_{\text{ent.}}^{\text{res.}} & = & +\lambda\frac{A_{\perp}}{12}\int\frac{d^{2}p_{\perp}}{(2\pi)^{2}}\frac{m_{\sigma}^{2}}{p_{\perp}^{2}+m_{\sigma}^{2}}\left.\left[D_{\sigma}(p_{\perp}^{2},q^{2})+(N-1)D_{\pi}(p_{\perp}^{2},q^{2})\right]\right|_{q^{2}=m_{\sigma}^{2}}\\
 &  & +\lambda\frac{(N-1)A_{\perp}}{12}\int\frac{d^{2}p_{\perp}}{(2\pi)^{2}}\frac{m_{\sigma}^{2}}{p_{\perp}^{2}}\left.\left[D_{\pi\sigma}(p_{\perp}^{2},q^{2})\right]\right|_{q^{2}=0}+O(\lambda^{2})\\
 & = & \lambda\frac{A_{\perp}}{12}\int\frac{d^{2}p_{\perp}}{(2\pi)^{2}}\frac{1}{(4\pi)^{2}}\left\{ \frac{m_{\sigma}^{2}}{p_{\perp}^{2}+m_{\sigma}^{2}}\left[9\int_{0}^{1}dx\ln\left(\frac{m_{\sigma}^{2}+x(1-x)m_{\sigma}^{2}}{m_{\sigma}^{2}+x(1-x)p_{\perp}^{2}}\right)+(N-1)\ln\left(\frac{m_{\sigma}^{2}}{p_{\perp}^{2}}\right)\right]\right.\\
 &  & \left.+\frac{m_{\sigma}^{2}}{p_{\perp}^{2}}\left[2(N-1)\int_{0}^{1}dx\ln\left(\frac{m_{\sigma}^{2}}{m_{\sigma}^{2}+xp_{\perp}^{2}}\right)\right]\right\} +O(\lambda^{2})\\
 & = & \ \frac{A_{\perp}\Lambda^{2}}{48\pi}\left\{ \frac{\lambda}{(4\pi)^{2}}\int_{0}^{1}dt\left\{ \frac{\tilde{m}_{\sigma}^{2}}{t+\tilde{m}_{\sigma}^{2}}\left[9\int_{0}^{1}dx\ln\left(\frac{\tilde{m}_{\sigma}^{2}+x(1-x)\tilde{m}_{\sigma}^{2}}{\tilde{m}_{\sigma}^{2}+x(1-x)t}\right)+(N-1)\ln\left(\frac{\tilde{m}_{\sigma}^{2}}{t}\right)\right]\right.\right.\\
 &  & \left.\left.+\frac{\tilde{m}_{\sigma}^{2}}{t}\left[2(N-1)\int_{0}^{1}dx\ln\left(\frac{\tilde{m}_{\sigma}^{2}}{\tilde{m}_{\sigma}^{2}+xt}\right)\right]\right\} +O(\lambda^{2})\right\} \\
 & = & \ \frac{A_{\perp}\Lambda^{2}}{48\pi}\left\{ \frac{\lambda}{(4\pi)^{2}}\int_{0}^{1}dt\left\{ \frac{\tilde{m}_{\sigma}^{2}}{t+\tilde{m}_{\sigma}^{2}}\left[-18\frac{\text{arcth}\sqrt{\frac{t}{t+4\tilde{m}_{\sigma}^{2}}}}{\sqrt{\frac{t}{t+4\tilde{m}_{\sigma}^{2}}}}+9\sqrt{5}\ln\left(\frac{3+\sqrt{5}}{2}\right)+(N-1)\ln\left(\frac{\tilde{m}_{\sigma}^{2}}{t}\right)\right]\right.\right.\\
 &  & \left.\left.+\frac{\tilde{m}_{\sigma}^{2}}{t}2(N-1)\left[1+(1+\frac{\tilde{m}_{\sigma}^{2}}{t})\ln\tilde{m}_{\sigma}^{2}/(t+\tilde{m}_{\sigma}^{2})\right]\right\} +O(\lambda^{2})\right\} \\
 & = & \ \frac{A_{\perp}\Lambda^{2}}{48\pi}\left\{ \frac{\lambda}{(4\pi)^{2}}\left\{ -18\tilde{m}_{\sigma}^{2}\text{Sl}_{2}(\tilde{m}_{\sigma}^{2})+\left[9\sqrt{5}\ln\left(\frac{3+\sqrt{5}}{2}\right)\right]\tilde{m}_{\sigma}^{2}\ln\left(\frac{\tilde{m}_{\sigma}^{2}+1}{\tilde{m}_{\sigma}^{2}}\right)\right.\right.\\
 &  & \left.\left.+(N-1)\tilde{m}_{\sigma}^{2}\left[\ln\left(\frac{\tilde{m}_{\sigma}^{2}+1}{\tilde{m}_{\sigma}^{2}}\right)\ln(\tilde{m}_{\sigma}^{2})-\text{Li}_{2}\left(-\frac{1}{\tilde{m}_{\sigma}^{2}}\right)\right]\right.\right.\\
 &  & \left.\left.+2(N-1)\tilde{m}_{\sigma}^{2}\left[-1+(1+\tilde{m}_{\sigma}^{2})\ln\left(\frac{\tilde{m}_{\sigma}^{2}+1}{\tilde{m}_{\sigma}^{2}}\right)+\text{Li}_{2}\left(-\frac{1}{\tilde{m}_{\sigma}^{2}}\right)\right]\right\} +O(\lambda^{2})\right\} \\
 & = & \ \frac{A_{\perp}\Lambda^{2}}{48\pi}\left\{ \frac{\lambda}{(4\pi)^{2}}\left\{ -\frac{9}{2}\tilde{m}_{\sigma}^{2}\ln^{2}(\tilde{m}_{\sigma}^{2})+O(\tilde{m}_{\sigma}^{2})-\left[9\sqrt{5}\ln\left(\frac{3+\sqrt{5}}{2}\right)\right]\tilde{m}_{\sigma}^{2}\ln(\tilde{m}_{\sigma}^{2})+O(\tilde{m}_{\sigma}^{2})\right.\right.\\
 &  & \left.\left.+(N-1)\tilde{m}_{\sigma}^{2}\left[-\ln^{2}(\tilde{m}_{\sigma}^{2})+\frac{1}{2}\text{ln}^{2}(\tilde{m}_{\sigma}^{2})+O(\tilde{m}_{\sigma}^{2})\right]\right.\right.\\
 &  & \left.\left.+2(N-1)\tilde{m}_{\sigma}^{2}\left[-\ln(\tilde{m}_{\sigma}^{2})-\frac{1}{2}\ln^{2}(\tilde{m}_{\sigma}^{2})+O(\tilde{m}_{\sigma}^{2})\right]\right\} +O(\lambda^{2})\right\} \\
 & = & \ \frac{A_{\perp}\Lambda^{2}}{48\pi}\left\{ \frac{\lambda}{(4\pi)^{2}}\left\{ -\left[\frac{9}{2}+\frac{3}{2}(N-1)\right]\tilde{m}_{\sigma}^{2}\ln^{2}(\tilde{m}_{\sigma}^{2})\right.\right.\\
 &  & \left.\left.-\left[9\sqrt{5}\ln\left(\frac{3+\sqrt{5}}{2}\right)+2(N-1)\right]\tilde{m}_{\sigma}^{2}\ln(\tilde{m}_{\sigma}^{2})\right\} +O(\lambda^{2},\tilde{m}_{\sigma}^{2})\right\} ,
\end{eqnarray*}
where $\text{Li}_{2}$ is the polylog function $\text{Li}_{\alpha=2}$,
and we also define the function of integral for the $\sigma$ loop
by 
\begin{eqnarray}
\text{Sl}_{2}(\tilde{m}_{\sigma}^{2}) & = & \int_{0}^{1}dt\frac{1}{t+\tilde{m}_{\sigma}^{2}}\left[\frac{\text{arcth}\sqrt{\frac{t}{t+4\tilde{m}_{\sigma}^{2}}}}{\sqrt{\frac{t}{t+4\tilde{m}_{\sigma}^{2}}}}\right],
\end{eqnarray}
such that 
\begin{eqnarray}
\lim_{x\rightarrow0}\frac{\text{Sl}_{2}(x)}{\ln^{2}x} & = & \frac{1}{4},\\
\lim_{x\rightarrow0}\frac{-\text{Li}_{2}(-\frac{1}{x})}{\ln^{2}x} & = & \frac{1}{2}.
\end{eqnarray}

To summarize, 
\begin{eqnarray}
\Delta S_{\text{ent.}}^{\text{res.}} & = & +\lambda\frac{A_{\perp}}{12}\int\frac{d^{2}p_{\perp}}{(2\pi)^{2}}\frac{m_{\sigma}^{2}}{p_{\perp}^{2}+m_{\sigma}^{2}}\left.\left[D_{\sigma}(p_{\perp}^{2},q^{2})+(N-1)D_{\pi}(p_{\perp}^{2},q^{2})\right]\right|_{q^{2}=m_{\sigma}^{2}}\nonumber \\
 &  & +\lambda\frac{(N-1)A_{\perp}}{12}\int\frac{d^{2}p_{\perp}}{(2\pi)^{2}}\frac{m_{\sigma}^{2}}{p_{\perp}^{2}}\left.\left[D_{\pi\sigma}(p_{\perp}^{2},q^{2})\right]\right|_{q^{2}=0}+O(\lambda^{2})\nonumber \\
 & = & \ \frac{A_{\perp}\Lambda^{2}}{48\pi}\left\{ \frac{\lambda}{(4\pi)^{2}}\int_{0}^{1}dt\left\{ \frac{\tilde{m}_{\sigma}^{2}}{t+\tilde{m}_{\sigma}^{2}}\left[9\int_{0}^{1}dx\ln\left(\frac{\tilde{m}_{\sigma}^{2}+x(1-x)\tilde{m}_{\sigma}^{2}}{\tilde{m}_{\sigma}^{2}+x(1-x)t}\right)+(N-1)\ln\left(\frac{\tilde{m}_{\sigma}^{2}}{t}\right)\right]\right.\right.\nonumber \\
 &  & \left.\left.+\frac{\tilde{m}_{\sigma}^{2}}{t}\left[2(N-1)\int_{0}^{1}dx\ln\left(\frac{\tilde{m}_{\sigma}^{2}}{\tilde{m}_{\sigma}^{2}+xt}\right)\right]\right\} +O(\lambda^{2})\right\} \nonumber \\
 & = & \ \frac{A_{\perp}\Lambda^{2}}{48\pi}\left[c_{N}^{'}\lambda\left(\frac{m_{\sigma}^{2}}{\Lambda^{2}}\right)\ln^{2}\left(\frac{m_{\sigma}^{2}}{\Lambda^{2}}\right)+c_{N}\lambda\left(\frac{m_{\sigma}^{2}}{\Lambda^{2}}\right)\ln\left(\frac{m_{\sigma}^{2}}{\Lambda^{2}}\right)+O\Big(\lambda^{2},\frac{m_{\sigma}^{2}}{\Lambda^{2}}\Big)\right],
\end{eqnarray}
where 
\begin{eqnarray}
c_{N} & = & -\frac{1}{(4\pi)^{2}}\left[\beta+2(N-1)\right],\\
c_{N}^{'} & = & -\frac{1}{(4\pi)^{2}}\left[\beta^{'}+\frac{3}{2}(N-1)\right],
\end{eqnarray}
with 
\begin{eqnarray}
\beta & = & 9\sqrt{5}\ln\frac{3+\sqrt{5}}{2},\\
\beta^{'} & = & 18\lim_{x\rightarrow0}\frac{1}{\ln^{2}x}\int_{0}^{1}dt\frac{1}{t+x}\frac{\text{arcth}\sqrt{\frac{t}{t+4x}}}{\sqrt{\frac{t}{t+4x}}}=\frac{9}{2}.
\end{eqnarray}
At last, we can see that there is an extra order that is more divergent
than $m_{\sigma}^{2}\log(m_{\sigma}^{2}/\Lambda^{2})$ but less divergent
than $1$ as $m_{\sigma}^{2}/\Lambda^{2}\rightarrow0$. This order
comes from all the two-vertex loops due to the cubic interactions.

  \bibliographystyle{jhep}
\addcontentsline{toc}{section}{\refname}\nocite{*}
\bibliography{ref_ee}

\end{document}